\documentclass[epsfig,onecolumn]{mn2e}
\usepackage{epsfig}
\def\beq#1{\begin{equation}\label{#1}}
\def\eeq{\end{equation}}
\def\beqa#1{\begin{eqnarray}\label{#1}}
\def\eeqa{\end{eqnarray}}

\def\h0{H$_{\rm 0}$~}

\title[High redshift cosmography and dark energy]{High redshift cosmography: new results and implication for dark energy  } \author[M.Demianski et al.]{Marek Demianski$^{1,2}$, Ester Piedipalumbo$^{3,4}$, Claudio Rubano$^{4}$, Paolo Scudellaro$^{3,4}$\\
$^1$ Institute for Theoretical Physics,
University of Warsaw,  Hoza 69, 00-681 Warsaw, Poland  \\
$^2$ Department of Astronomy, Williams College, Williamstown, MA 01267, USA\\
$^3$ Dipartimento di Scienze Fisiche, Universit\`{a} di Napoli
Federico II, Compl.
Univ. Monte S. Angelo, 80126 Naples, Italy  \\
$^4$ I.N.F.N., Sez. di Napoli, Complesso Universitario di Monte
Sant' Angelo, Edificio G, via Cinthia, 80126 - Napoli, Italy }
\date{Accepted xxx, Received yyy, in original form zzz}
\begin{document}
\maketitle
\begin{abstract}
The explanation of the accelerated expansion of the Universe poses one of
the most fundamental questions in physics and cosmology today. If the acceleration is driven by
some form of dark energy, and in the absence of a well-based theory to interpret the observations,
one can try to constrain the parameters describing the kinematical state
of the universe using a cosmographic approach, which is fundamental in that it requires only a minimal set of assumptions, namely to specify the metric, and it does not rely on the dynamical equations for gravity. Our high-redshift analysis allows us to put constraints on the cosmographic expansion up to the fifth order. It is based on the Union2 Type Ia Supernovae (SNIa) data set, the Hubble diagram constructed from some Gamma Ray Bursts luminosity distance indicators, and gaussian priors on the distance from the Baryon Acoustic Oscillations (BAO), and the Hubble constant $h$ (these priors have been included in order to help break the degeneracies among model parameters). To perform our statistical analysis and to explore the probability distributions of the cosmographic parameters we use the Markov Chain Monte Carlo Method (MCMC). We finally investigate implications of our results for the dark energy, in particular, we focus on the parametrization of the dark energy equation of state (EOS). Actually, a possibility to investigate the nature of dark energy lies in measuring the dark energy equation of state, $w$, and its time (or redshift) dependence at high accuracy. However, since $w(z)$ is not directly accessible
to measurement, reconstruction methods are needed to extract it reliably from observations.
Here we investigate different models of dark energy, described through several
parametrizations of the equation of state, by comparing the cosmographic and the EOS series.
The main results are: $a)$ even if relying on a mathematical approximate assumption
such as the scale factor series expansion in terms of time,
cosmography can be extremely useful in assessing
dynamical properties of the Universe; $b)$ the deceleration parameter
clearly confirms the present acceleration phase; $c)$ the MCMC method
provides stronger constraints for parameter estimation, in
particular for higher order cosmographic parameters (the jerk and
the snap), with respect to those presented in the literature; $d
)$ both the
estimation of the jerk and the DE parameters, reflect the
possibility of a deviation from the $\Lambda$CDM cosmological
model; $e)$ there are indications that the dark energy equation  of state is evolving for all the parametrizations that we considered; $f)$ the $q(z)$ reconstruction provided by our cosmographic analysis allows a transient acceleration.
\end{abstract}
\begin{keywords}
Gamma Rays\,: bursts -- Cosmology\,: distance scale -- Cosmology\,:
cosmological parameters
\end{keywords}
\section{Introduction}
At the end  of the '90s observations of high redshift supernovae of type Ia (SNIa) revealed
that the universe is now expanding at an accelerated rate. This
surprising result has been independently confirmed by observations
of small scale temperature anisotropies of the cosmic microwave
background radiation (CMB) \cite{Riess07,SNLS,Union,WMAP3}. It is
usually assumed that the observed accelerated expansion is caused by
a so called dark energy, with unusual properties. The pressure of
dark energy $p_{de}$ is negative and it is related to the positive
energy density of dark energy $\epsilon_{de}$ by
$p_{de}=w\epsilon_{de}$ where the proportionality coefficient $w<0$.
According to the present day estimates, about 75\% of matter-energy
in the universe is in the form of dark energy, so that now the dark
energy is the dominating component in the universe. The nature of
dark energy is not known. Proposed so far models of dark energy can
be divided, at least, into three groups: a) a non zero cosmological
constant, in this case $w=-1$,  b) a potential energy of some not
yet discovered scalar field, or c) effects connected with non
homogeneous distribution of matter and averaging procedures.
In most scenarios of dark energy, its properties are mainly characterized by the
equation of state parameter (EOS), $w$. Extracting the information
on EOS of dark energy from observational data is then at the same time a fundamental problem and a challenging task. For probing the dynamical evolution of dark energy, under such circumstance, one can parameterize $w$ empirically, usually using two or more free parameters. Among all the parametrization forms of EOS,  we will consider
the Chevallier-Polarski-Linder (CPL) model~\cite{cpl1,cpl2}, which is probably the most widely used parametrization, since it presents  a smooth and bounded
behavior for high redshifts, and a manageable two-dimensional parameter space. However we will also consider novel parametrizations recently introduced in \cite{novel} and \cite{oscisalz} to avoid the future divergency problem in the CPL
parametrization, which turned out to be able to genuinely cover many scalar-field models, as well as other theoretical scenarios.
It is worth noticing that all the dark energy models considered so far are in agreement with the observational data. As a consequence, unless higher precision probes of the expansion rate and the growth of structure are developed, these different approaches cannot be discriminated. This degeneration suggests a \textit{kinematical} approach to the problem of cosmic acceleration, relying on quantities that are only weakly model dependent.
The cosmographic approach is only related to the derivatives of the scale factor and it makes it possible to fit the data on the distance - redshift relation without any a priori assumption on the underlying cosmological model. It is based on the only assumption that the metric is spatially homogeneous and isotropic.
The SNIa Hubble diagram extends up to $z = 1.7$ thus invoking the need for, at least, a fifth order Taylor expansion of the scale factor in order to give a reliable approximation of the distance - redshift relation. As a consequence, it could be, in principle, possible to estimate up to five cosmographic parameters, $(h, q_0, j_0, s_0, l_0)$ , although the still too small data set available does not allow to get a precise and realistic determination of all of them.
Once these quantities have been determined, one could use them to put constraints on the dark energy models. We are reverting the usual approach that attempts to derive the cosmographic parameters as a sort of byproduct of the assumed theory. Here, we use the cosmographic parameters to parametrize the quantities that  characterize the model so that each dark energy model is characterized by the same set of parameters $(h, q_0, j_0, s_0, l_0)$.
For constraining the cosmographic parameters, we use the Union2 Type Ia Supernovae (SNIa) data set, the Hubble diagram constructed from some Gamma Ray Bursts luminosity distance indicators, and gaussian priors on the distance from the Baryon Acoustic Oscillations (BAO), and the Hubble constant $h$ (such priors have been included in order to help break the degeneracies among model parameters). Actually, observations of the type Ia supernovae are consistent with the assumption that the observed accelerated expansion is due to the non zero cosmological constant. However, so far the type Ia supernovae have been observed only at redshifts $z<2$, while in order to test if $w$ is changing with redshift it is necessary to use more distant objects. New possibilities opened up when  the Gamma Ray Bursts have been discovered at  higher redshifts, the present record is at $z=8.26$ \cite{Greiner08}.
GRBs are however  enigmatic objects. First of all the mechanism that is responsible for releasing the incredible amounts of energy that a typical GRB emits is not yet known (see for instance Meszaros 2006 for a recent review). It is also not yet definitely known if the energy is emitted isotropically or is beamed. Despite
of these difficulties GRBs are promising objects that can be used to study the expansion rate of the universe at high redshifts
\cite{Bradley03,S03,Dai04,Bl03,Firmani05,S07,Li08,Amati08,Ts09}. Actually even if the huge dispersion (about four orders of magnitude) of the isotropic GRB energy makes them everything but standard candles, it has been recently empirically established that some of the directly observed parameters of GRBs are correlated with their important intrinsic parameters, like the luminosity or the total radiated energy, allowing to derive  some correlations,
which have been tested and used to standardize GRBs and to calibrate these relations, and to derive
their luminosity or radiated energy from one or more observables, in order to construct a GRBs Hubble diagram. It has been shown that such procedure can be implemented without specifying the cosmological model, see for instance, \cite{MEC10,ME} and references therein. In our analysis we use two GRB HD data sets: one sample consists of $109$ high redshift GRBs and has been constructed from the
Amati $E_{\rm p,i}$ -- $E_{\rm iso}$ correlation (here $E_{\rm p,i}$ is the peak photon energy of the intrinsic
spectrum and $E_{\rm iso}$ the isotropic equivalent radiated energy), applying a local regression technique to estimate, in a model independent way,
the distance modulus from the recently updated Union SNIa data set. The second GRBs HD sample is constructed from 66 Gamma Ray Bursts (GRBs) derived using only data from their X\,-\,ray afterglow light curve. To this end, we used the recently updated $L_X$\,-\,$T_a$ correlation between the break time $T_a$ and the X\,-\,ray luminosity $L_X$ measured at $T_a$ calibrated (using SNIa) from a sample of {\it Swift} GRBs \cite{vinclxta}.
It is worth noting that such GRBs HD are based on the use of a single correlation and contain a statistically meaningful number of objects.
The use of the Amati, $E_{\rm p,i}$ -- $E_{\rm iso}$, and the $L_X$\,-\,$T_a$ correlations then avoids the need of combining different correlations
to increase the number of GRBs with a known distance modulus. Since each correlation is affected by its own possible systematics and characterized by different intrinsic scatter so that combining all of them in a single HD can introduce unexpected features and hence bias the constraints on the cosmology.
It turns out that these data sets are sufficient for our aim of testing and comparing the new parametrizations. Moreover in order to check if the results of our cosmographic analysis are  biased due to the procedure used to calibrate the GRBs correlations we performed a consistency test: we actually apply a full bayesian approach, extracting, at the same time, the correlation coefficients and the cosmological parameters of the model from the observed quantities.
Also to accomplish this task, we use the Markov Chain Monte Carlo  simulations and compute, simultaneously, the full probability density functions (PDFs) of all the parameters of interest.
This approach does not require any prior information on the cosmological model and yields results that are not plagued by any of the various limitations known in the literature, see for instance \cite{MEC10,ME}. Since such a procedure is a demanding job from the point of view of computation time, it has been applied for the Amati relation only. The results turned out to be fully statistically consistent with the ones obtained by performing a local regression technique, thus indicating that it is not affected by any systematic bias induced by the calibration procedure.
For the other data set we use the Markov Chain Monte Carlo  simulations just to perform the cosmological tests.

The scheme of the paper is as follows. In Section 2 we describe the basic elements of the cosmographic approach and explicitly derive series expansions of the scale factor and other relevant parameters. In Section 3 we describe the observational data sets that are used in our analysis. In Section 4 we describe some details of our statistical analysis and present results on cosmographic parameters obtained from three sets of data. In Section 5 we present constrains on dark energy models that can be derived from our analysis. General discussion of our results and conclusions are presented in Section 6.

\section{The cosmography approach }
Recently the cosmographic approach to cosmology  gained increasing interest for catching as much information
as possible directly from observations, retaining the minimal priors
of isotropy and homogeneity and leaving aside other
assumptions. Actually, the only ingredient taken into
account \textit{a priori} in this approach  is the FLRW line element obtained from
kinematical requirements
\begin{equation}
ds^2=-c^2dt^2+a^2(t)\left[\frac{dr^2}{1-kr^2}+r^2d\Omega^2\right]\,,
\end{equation}
where $a(t)$ is the scale factor and $k= +1, 0, - 1$ is the curvature parameter.
Using this metric, it is possible to express the luminosity
distance $d_L$ as a power series in the redshift parameter $z$, the
coefficients of the expansion being functions of the scale factor
$a(t)$ and its higher order derivatives. This
expansion leads to a distance\,-\,redshift relation which only
relies on the assumption of the FLRW metric thus
being fully model independent since it does not depend on the
particular form of the solution of cosmic evolution equations. To this aim,
it is convenient to introduce the following cosmographic functions \cite{Visser}:
\begin{eqnarray}\label{eq:cosmopar}
H(t) &\equiv& + \frac{1}{a}\frac{da}{dt}\, ,
\\
q(t) &\equiv& - \frac{1}{a}\frac{d^{2}a}{dt^{2}}\frac{1}{H^{2}}\,
,
\\
j(t) &\equiv& + \frac{1}{a}\frac{d^{3}a}{dt^{3}}\frac{1}{H^{3}}\,
,
\\
s(t) &\equiv& + \frac{1}{a}\frac{d^{4}a}{dt^{4}}\frac{1}{H^{4}}\,
,
\\
l(t) &\equiv& + \frac{1}{a}\frac{d^{5}a}{dt^{5}}\frac{1}{H^{5}}\,
.
\end{eqnarray}
When evaluated at the present time $t_0$ these functions correspond to the cosmographic parameters, which are usually referred to as the \textit{Hubble},
\textit{deceleration}, \textit{jerk}, \textit{snap} and
\textit{lerk} parameters, respectively\footnote{Note that the
use of the jerk parameter to discriminate between different models was also
proposed in \cite{SF} in the context of the {\it statefinder}
parametrization.}. Furthermore, it is possible
to relate the derivative of the Hubble parameter to the other
cosmographic parameters\,:
{\setlength\arraycolsep{0.2pt}
\begin{eqnarray}
\dot{H} &=& -H^2 (1 + q) \ , \label{eq: hdot}
\\
\ddot{H} &=& H^3 (j + 3q + 2) \ , \label{eq: h2dot}
\\
d^3H/dt^3 &=& H^4 \left ( s - 4j - 3q (q + 4) - 6 \right ) \ ,
\label{eq: h3dot}
\\
d^4H/dt^4 &=& H^5 \left ( l - 5s + 10 (q + 2) j + 30 (q + 2) q +
24 \right ) \ , \label{eq: h4dot}
\end{eqnarray}}
where a dot denotes derivative with respect to the cosmic time
$t$. With these definitions the series expansion to the 5th order in
time of the scale factor is: {\setlength\arraycolsep{0.2pt}
%%\begin{eqnarray}
%%a(t) &=& a(t_{0}) \left\{1 + H_{0} (t-t_{0}) - \frac{q_{0}}{2}
%%H_{0}^{2} (t-t_{0})^{2} +  \frac{j_{0}}{3!} H_{0}^{3}
%%(t-t_{0})^{3} + \frac{s_{0}}{4!} H_{0}^{4} (t-t_{0})^{4} +
%%\frac{l_{0}}{5!} H_{0}^{5} (t-t_{0})^{5}
%%+\emph{O}[(t-t_{0})^{6}]\right\}\,,
%%\end{eqnarray}}
{\setlength\arraycolsep{0.2pt}
\begin{eqnarray}\label{eq:a_series}
\frac{a(t)}{a(t_{0})} &=& 1 + H_{0} (t-t_{0}) -\frac{q_{0}}{2}
H_{0}^{2} (t-t_{0})^{2} +\frac{j_{0}}{3!} H_{0}^{3} (t-t_{0})^{3}
+ \frac{s_{0}}{4!} H_{0}^{4} (t-t_{0})^{4}+ \frac{l_{0}}{5!}
H_{0}^{5} (t-t_{0})^{5} +\emph{O}[(t-t_{0})^{6}]\,.
\end{eqnarray}}
From Eq.(\ref{eq:a_series}), and  remembering that
the distance traveled by a photon that is emitted at
time $t_{*}$ and absorbed at the current epoch $t_{0}$ is
\begin{equation}
D = c \int dt = c (t_{0} - t_{*})\,,
\end{equation}
we can construct the series for $z(D)$, actually
\begin{eqnarray}
z(D)
 &=& \mathcal{Z}_{D}^{1} \left(\frac{H_{0} D}{c}\right) +
\mathcal{Z}_{D}^{2} \left(\frac{H_{0} D}{c}\right)^{2} +
\mathcal{Z}_{D}^{3} \left(\frac{H_{0} D}{c}\right)^{3} +
\mathcal{Z}_{D}^{4} \left(\frac{H_{0} D}{c}\right)^{4} +
\mathcal{Z}_{D}^{5} \left(\frac{H_{0} D}{c}\right)^{5} +
\emph{O}\left[\left(\frac{H_{0} D}{c}\right)^{6}\right]\,,\label{zdseries}
\end{eqnarray}
with: {\setlength\arraycolsep{0.2pt}
\begin{eqnarray}
\mathcal{Z}_{D}^{1} &=& 1\,, \\
\mathcal{Z}_{D}^{2} &=& 1 + \frac{q_{0}}{2}\,, \\
\mathcal{Z}_{D}^{3} &=& 1 + q_{0} +\frac{j_{0}}{6}\,, \\
\mathcal{Z}_{D}^{4} &=& 1 + \frac{3}{2}q_{0} + \frac{q_{0}^{2}}{4} + \frac{j_{0}}{3} - \frac{s_{0}}{24}\,, \\
\mathcal{Z}_{D}^{5} &=& 1 + 2 q_{0} + \frac{3}{4} q_{0}^{2} +
\frac{q_{0} j_{0}}{6} + \frac{j_{0}}{2} - \frac{s}{12} + l_{0}\,.
\end{eqnarray}}

To obtain the
physical distance $D$ expressed as a function of redshift $z$ we reverse the series $z(D) \rightarrow D(z)$, obtaining:

\begin{eqnarray}
D(z) &=& \frac{c z}{H_{0}} \left( \mathcal{D}_{z}^{0} +
\mathcal{D}_{z}^{1}  z + \mathcal{D}_{z}^{2}  z^{2} +
\mathcal{D}_{z}^{3}  z^{3} + \mathcal{D}_{z}^{4}  z^{4} +
\emph{O}(z^{5}) \right)\,,
\end{eqnarray}
with: {\setlength\arraycolsep{0.2pt}
\begin{eqnarray}
\mathcal{D}_{z}^{0} &=& 1\,, \\
\mathcal{D}_{z}^{1} &=& - \left(1 +\frac{ q_{0}}{2}\right)\,, \\
\mathcal{D}_{z}^{2} &=& 1 + q_{0} + \frac{q_{0}^{2}}{2} - \frac{j_{0}}{6}\,, \\
\mathcal{D}_{z}^{3} &=& - \left(1 + \frac{3}{2}q_{0}+
\frac{3}{2}q_{0}^{2} + \frac{5}{8} q_{0}^{3} - \frac{1}{2} j_{0} -
\frac{5}{12} q_{0} j_{0} - \frac{s_{0}}{24}\right)\,, \\
\mathcal{D}_{z}^{4} &=& 1 + 2 q_{0} + 3 q_{0}^{2} + \frac{5}{2}
q_{0}^{3} + \frac{7}{2} q_{0}^{4} - \frac{5}{3} q_{0}
j_{0} - \frac{7}{8} q_{0}^{2} j_{0} - \frac{1}{8} q_{0} s_{0} - j_{0} +\frac{j_{0}^{2}}{12} - \frac{s_{0}}{6} - \frac{l_{0}}{120}\,.
\end{eqnarray}}

In the following we will be interested not in the physical
distance $D(z)$, but in the luminosity or angular-diameter distance, which can be calculated as
\begin{equation}
    d_{L} = \frac{a(t_{0})}{a(t_{0}-\frac{D}{c})} \: (a(t_{0}) r_{0})\,,
    \end{equation}

    \begin{equation}
    d_{A} = \frac{a(t_{0}-\frac{D}{c})}{a(t_{0})} \: (a(t_{0}) r_{0})\,,
    \end{equation}
where $r_{0}(D)$ is:
\begin{equation}\label{eq:r_sin}
r_{0}(D) = \left\{
\begin{array}{lr}
  \sin ( \int_{t_{0}- \frac{D}{c}}^{t_{0}} \frac{c \ \mathrm{d}t}{a(t)} ) &  k = +1; \\
  &  \\
  \int_{t_{0}- \frac{D}{c}}^{t_{0}} \frac{c \ \mathrm{d}t}{a(t)} &  k = 0; \\
  &  \\
  \sinh ( \int_{t_{0}- \frac{D}{c}}^{t_{0}} \frac{c \ \mathrm{d}t}{a(t)} ) &  k = -1.
\end{array} \right.
\end{equation}
If  we insert
the series expansion of $a(t)$ in $r_{0}(D)$, we have the cosmographic expansion of $r_{0}(D)$:
{\setlength\arraycolsep{0.2pt}
\begin{eqnarray}
r_{0}(D) &=& \int_{t_{0} - \frac{D}{c}}^{t_{0}} \frac{c \
\mathrm{d}t}{a(t)} = \int_{t_{0} - \frac{D}{c}}^{t_{0}} \frac{c \
\mathrm{d}t}{a_{0}} \left\{ 1 + H_{0} (t_{0} - t) + \left(1 +
\frac{q_{0}}{2}\right) H_{0}^{2}(t_{0} - t)^{2} + \left(1 + q_{0}
+\frac{ j_{0}}{6}\right)H_{0}^{3}(t_{0} - t)^{3} + \right. \nonumber \\
&+& \left. \left(1 + \frac{3}{2}q_{0} + \frac{q_{0}^{2}}{4} +
\frac{j_{0}}{3} - \frac{s_{0}}{24}\right)H_{0}^{4}(t_{0} - t)^{4}
+ \left(1 + 2 q_{0} + \frac{3}{4} q_{0}^{2} + \frac{q_{0}
j_{0}}{6} + \frac{j_{0}}{2} - \frac{s}{12} +
l_{0}\right)H_{0}^{5}(t_{0} - t)^{5} + \emph{O}[(t_{0} - t)^{6}]
\right\} = \nonumber \\ &=& \frac{D}{a_{0}} \left\{ 1 +
\frac{1}{2} \frac{H_{0} D}{c} + \left[\frac{2 + q_{0}}{6}\right]
\left(\frac{H_{0} D}{c}\right)^{2} + \left[ \frac{6 + 6 q_{0} +
j_{0}}{24} \right] \left(\frac{H_{0} D}{c}\right)^{3} +
\left[ \frac{24 + 36 q_{0} + 6 q_{0}^{2} + 8 j_{0} - s_{0}}{120} \right] \left(\frac{H_{0} D}{c}\right)^{4} + \right. \nonumber \\
&+& \left. \left[ \frac{12 + 24 q_{0} + 9 q_{0}^{2} + 2 q_{0}
j_{0} + 6 j_{0} - s_{0} + 12 l_{0}}{72} \right] \left(\frac{H_{0}
D}{c}\right)^{5}+ \emph{O}\left[\left(\frac{H_{0}
D}{c}\right)^{6}\right] \right\}\,.
\end{eqnarray}}
In  our analysis we will consider spatially flat cosmological models only, so that
\begin{eqnarray}
r_{0}(D) &=& \frac{D}{a_{0}} \left\{ \mathcal{R}_{D}^{0} +
\mathcal{R}_{D}^{1} \frac{H_{0} D}{c} + \mathcal{R}_{D}^{2}
\left(\frac{H_{0} D}{c}\right)^{2} + \mathcal{R}_{D}^{3}
\left(\frac{H_{0} D}{c}\right)^{3} + \right. \nonumber \\
&+&\left. \mathcal{R}_{D}^{4} \left(\frac{H_{0} D}{c}\right)^{4} +
\mathcal{R}_{D}^{5} \left(\frac{H_{0} D}{c}\right)^{5} +
\emph{O}\left[\left(\frac{H_{0} D}{c}\right)^{6}\right] \right\}\,,
\end{eqnarray}
with: {\setlength\arraycolsep{0.2pt}
\begin{eqnarray}
\mathcal{R}_{D}^{0} &=& 1\,, \\
\mathcal{R}_{D}^{1} &=& \frac{1}{2}\,, \\
\mathcal{R}_{D}^{2} &=& \frac{1}{6} \left(2 + q_{0} \right)\,, \\
\mathcal{R}_{D}^{3} &=& \frac{1}{24} \left( 6 + 6 q_{0} + j_{0} -\right)\,,\\
\mathcal{R}_{D}^{4} &=& \frac{1}{120} \left( 24 + 36 q_{0} + 6 q_{0}^{2} + 8 j_{0} - s_{0} \right)\,, \\
\mathcal{R}_{D}^{5} &=& \frac{1}{144} \left( 24 + 48 q_{0} + 18
q_{0}^{2} + 4 q_{0} j_{0} + 12 j_{0} - 2 s_{0} + 24 l_{0} \right)\,.
\end{eqnarray}}
With this expansion the luminosity distance is given as:
\begin{eqnarray}\label{serielum1}
d_{L}(z) = \frac{c z}{H_{0}} \left( \mathcal{D}_{L}^{0} +
\mathcal{D}_{L}^{1} \ z + \mathcal{D}_{L}^{2} \ z^{2} +
\mathcal{D}_{L}^{3} \ z^{3} + \mathcal{D}_{L}^{4} \ z^{4} +
\emph{O}(z^{5}) \right)\,,
\end{eqnarray}
with: {\setlength\arraycolsep{0.2pt}
\begin{eqnarray}\label{serieslum2}
\mathcal{D}_{L}^{0} &=& 1\,, \\
\mathcal{D}_{L}^{1} &=& - \frac{1}{2} \left(-1 + q_{0}\right)\,, \\
\mathcal{D}_{L}^{2} &=& - \frac{1}{6} \left(1 - q_{0} - 3q_{0}^{2} + j_{0}\right)\,, \\
\mathcal{D}_{L}^{3} &=& \frac{1}{24} \left(2 - 2 q_{0} - 15
q_{0}^{2} - 15 q_{0}^{3} + 5 j_{0} + 10 q_{0} j_{0} + s_{0} \right)\,,\\
\mathcal{D}_{L}^{4} &=& \frac{1}{120} \left( -6 + 6 q_{0} + 81
q_{0}^{2} + 165 q_{0}^{3} + 105 q_{0}^{4} - 110 q_{0} j_{0} - 105
q_{0}^{2} j_{0} - 15 q_{0} s_{0} + \right.\nonumber \\
&-& \left.  27 j_{0} + 10 j^{2} - 11 s_{0} - l_{0}\right)\,.
\end{eqnarray}}
While for the angular diameter distance we get:
\begin{eqnarray}
d_{A}(z) = \frac{c z}{H_{0}} \left ( \mathcal{D}_{A}^{0} +
\mathcal{D}_{A}^{1}  z + \mathcal{D}_{A}^{2} \ z^{2} +
\mathcal{D}_{A}^{3}  z^{3} + \mathcal{D}_{A}^{4}  z^{4} +
\emph{O}(z^{5}) \right )\,,
\end{eqnarray}
with: {\setlength\arraycolsep{0.2pt}
\begin{eqnarray}
\mathcal{D}_{A}^{0} &=& 1 \,,\\
\mathcal{D}_{A}^{1} &=& - \frac{1}{2} \left(3 + q_{0}\right)\,, \\
\mathcal{D}_{A}^{2} &=& \frac{1}{6} \left(11 + 7 q_{0} + 3q_{0}^{2} - j_{0} \right)\,, \\
\mathcal{D}_{A}^{3} &=& - \frac{1}{24} \left(50 + 46 q_{0} + 39
q_{0}^{2} + 15 q_{0}^{3} - 13 j_{0} - 10 q_{0} j_{0} - s_{0}
- \frac{2 k c^{2} (5 + 3 q_{0})}{H_{0}^{2} a_{0}^{2}}\right)\,, \\
\mathcal{D}_{A}^{4} &=& \frac{1}{120} \left( 274 + 326 q_{0} + 411
q_{0}^{2} + 315 q_{0}^{3} + 105 q_{0}^{4} - 210 q_{0} j_{0} - 105
q_{0}^{2} j_{0} - 15 q_{0} s_{0} + \right. \\
&-& \left. 137 j_{0} + 10 j^{2} - 21 s_{0} - l_{0} \right)\,.\nonumber
\end{eqnarray}}
It is worth noting that one can obtain the same
final expression for the  distance  starting from the Taylor
series expansion of the Hubble parameter instead of the scale factor,
namely:
{\setlength\arraycolsep{0.2pt}
\begin{eqnarray}\label{eq:Hseriesdef}
H(z) &=& H_{0} + \frac{dH}{dz}\Bigg{|}_{z=0} z + \frac{1}{2!}
\frac{d^{2}H}{dz^{2}}\Bigg{|}_{z=0} z^{2} + \frac{1}{3!}
\frac{d^{3}H}{dz^{3}}\Bigg{|}_{z=0} z^{3} +
\frac{1}{4!} \frac{d^{4}H}{dz^{4}}\Bigg{|}_{z=0} z^{4} +
\emph{O}(z^{5}) \, .
\end{eqnarray}}
To compute all the terms we use the derivation rule
\begin{equation}\label{eq:firstHt}
\frac{d}{dt} = -(1 + z) H \frac{d}{dz}\,.
\end{equation}
The series expansion of the Hubble parameter  will be used in our analysis in the
definition of our Markov chains algorithm given the observational data we use.
It is worth noting that since the cosmography is based on series expansions, the fundamental difficulties of applying such an approach to fit the luminosity distance data using high redshift distance indicators are connected with the
convergence and the truncation of the series. Recently the possibility of attenuating the convergence problem has been analyzed by defining a new redshift variable, see \cite{vitagliano}, the so called \textit{y-redshift}:
\begin{equation}
z \rightarrow y = \frac{z}{1+z}~.
\end{equation}
It turns out that for a series expansion in the classical z-redshift the
convergence radius is equal to $1$, which is a drawback
when one wants to extend the application of cosmography to
redshifts $z>1$.
The y-redshift could potentially weaken this problem because the z-interval
$[0,\infty]$ corresponds to the y-interval $[0,1]$, so that we are
mainly inside the convergence interval of the series, even for CMB
data ($z = 1089 \rightarrow y = 0.999$). Thus, in principle, we
could extend the series up to the redshift of decoupling, and
one could place CMB related constraints within the cosmographic
approach  \footnote{Let us note that the introduction of
this new redshift variable will not affect the definition of
cosmographic parameters.}.
However even using the series expansions in y-redshift the problem of the truncation of the series remains.
Here we consider a forth order expansion and are able to successfully put bounds on the parameters in a
statistically  consistent way.
In order to give reasonably narrow statistical constraints we apply a Markov Chain Monte Carlo
(MCMC) method, which allows us to obtain marginalized
likelihoods on the series coefficients from which we infer
rather tight constraints on those parameters. Actually, in our code we have inserted several tests, which give us control over
several physical requirements we expect from the theory.
For instance, since we use data related to the Hubble parameter $H(z)$, we are able to set restrictions on the Hubble parameter, $H_0=H(0)$, and thus to obtain a considerable improvement in the quality of constraints.

\section{Observational data sets}
\label{s3}
In our cosmographic approach we use  the currently available observational data sets on SNIa and GRB Hubble Diagrams, and we set gaussian priors on the distance from the Baryon Acoustic Oscillations (BAO), and the Hubble constant $h$. Such priors have been included in order to help break the degeneracies among the parameters of the cosmographic series expansion in Eqs. (\ref{serielum1}).
\subsection{Supernovae}
\label{sec:SNdata}
 Over the last decade the confidence in type Ia supernovae as standard candles has been steadily
growing.  Actually, the SNIa observations gave the first strong indication of an accelerating
expansion of the universe, which can be explained by assuming the existence of some kind of dark energy or
nonzero cosmological constant. Since 1995 two teams of astronomers - the High-Z
Supernova Search Team and the Supernova Cosmology Project - have been discovering type Ia supernovae at high
redshifts. First results of both teams were published by  \cite{Riess} and
\cite{per+al99}.
Here we consider the recently updated Supernovae Cosmology Project \textit{Union2} compilation
\cite{Union2}, which is an update of the original \textit{Union} compilation, now bringing together data on
$719$ SN, drawn from $17$ data sets. Of these, $557$ SN, spanning the redshift range ($0.015 \le z \le 1.55$),
pass usability cuts and outliers removal, and form the final sample used to constrain our model. We actually
compare the \textit{theoretically\thinspace\ predicted} distance modulus $\mu(z)$
with the \textit{observed} one, through a Bayesian approach, based on the definition
of the distance modulus,
\begin{equation}
\mu(z_{j}) = 5 \log_{10} ( D_{L}(z_{j}, \{\theta_{i}\}) )+\mu_0\,,
\end{equation}
where $D_{L}(z_{j}, \{\theta_{i}\})$ is the Hubble free luminosity
distance, expressed as a series depending on the cosmographic parameters, $\theta_{i}=(q_{0},
j_{0}, s_{0}, l_{0})$. The best fits were obtained by minimizing the
quantity
\begin{equation}\label{eq: sn_chi}
\chi^{2}_{\mathrm{SN}}(\mu_{0}, \{\theta_{i}\}) = \sum^{557}_{j =
1} \frac{(\mu(z_{j}; \mu_{0}, \{\theta_{i})\} -
\mu_{obs}(z_{j}))^{2}}{\sigma^{2}_{\mathrm{\mu},j}}\,,
\end{equation}
where the $\sigma^{2}_{\mathrm{\mu},j}$ are the measurement
variances. The parameter $\mu_{0}$ encodes the Hubble
constant and the absolute magnitude $M$, and has to be
marginalized over. Giving the heterogeneous origin of the Union data
set, and the procedures  for reducing data, we have worked with an alternative version of
Eq.~(\ref{eq: sn_chi}), which consists in minimizing the quantity
\begin{equation}\label{eq: sn_chi_mod}
\tilde{\chi}^{2}_{\mathrm{SN}}(\{\theta_{i}\}) = c_{1} -
\frac{c^{2}_{2}}{c_{3}}\,,
\end{equation}
with respect to the other parameters, where
\begin{equation}
c_{1} = \sum^{557}_{j = 1} \frac{(\mu(z_{j}; \mu_{0}=0,
\{\theta_{i})\} -
\mu_{obs}(z_{j}))^{2}}{\sigma^{2}_{\mathrm{\mu},j}}\, ,
\end{equation}
\begin{equation}
c_{2} = \sum^{557}_{j = 1} \frac{(\mu(z_{j}; \mu_{0}=0,
\{\theta_{i})\} -
\mu_{obs}(z_{j}))}{\sigma^{2}_{\mathrm{\mu},j}}\, ,
\end{equation}
\begin{equation}
c_{3} = \sum^{557}_{j = 1}
\frac{1}{\sigma^{2}_{\mathrm{\mu},j}}\,.
\end{equation}
It turns out that $\tilde{\chi}^{2}_{SN}$ is just a
version of $\chi^{2}_{SN}$, minimized with respect to $\mu_{0}$.
Actually, we find that
\begin{equation}
\chi^{2}_{\mathrm{SN}}(\mu_{0}, \{\theta_{i}\}) = c_{1} - 2 c_{2}
\mu_{0} + c_{3} \mu^{2}_{0} \,,
\end{equation}
which clearly becomes minimum for $\mu_{0} = c_{2}/c_{3}$, and so
we see that $\tilde{\chi}^{2}_{\mathrm{SN}} \equiv
\chi^{2}_{\mathrm{SN}}(\mu_{0} = c_{2}/c_{3}, \{\theta_{i}\})$. Furthermore,
one can check that the difference between $\chi^{2}_{SN}$ and
$\tilde{\chi}^{2}_{SN}$ is negligible.
\subsection{GRBs Hubble diagram}
By virtue of their enormous energy release, GRBs are visible up to very high $z$, and hence are ideal candidates for our high-redshift cosmography task. Unfortunately, GRBs are everything but standard candles because their peak luminosity spans a wide range. Nevertheless there have been many attempts to make them standardizeable candles resorting to the use of some empirical correlations among distance dependent quantities and rest frame observables \cite{Amati08}. Such empirical relations allow one to infer the GRB rest frame luminosity or energy from an observer frame measured quantity so that the distance modulus can be estimated with an error mainly depending on the intrinsic scatter of the adopted correlation. Combining the estimates from different correlations, Schaefer (2007) first derived the GRBs HD for 69 objects, which has been further enlarged using updated samples, different calibration methods and also different correlation relations, see for instance \cite{CCD}, \cite{MEC10}, \cite{ME}, showing the interest in the cosmological applications of GRBs. In this paper we perform our cosmographic analysis using two GRBs HD data set, build up by calibrating the Amati $E_{\rm p,i}$ -- $E_{\rm iso}$ and the $L_X$\,-\,$T_a$ correlations respectively.
\subsubsection{The \textit{calibrated Amati} Gamma Ray Bursts Hubble diagram}
It has been recently empirically established that some of the directly observed parameters of Gamma Ray Bursts
are connected with the isotropic absolute luminosity $L_{iso}$, the collimation corrected energy $E_{\gamma}$,
or the isotropic bolometric energy $E_{\rm iso}$ of a GRB.
These quantities appear to correlate with the GRB isotropic luminosity, its total collimation-corrected or its
isotropic energy. The isotropic luminosity cannot be measured directly but rather it can be obtained through the knowledge
of either the bolometric peak flux, denoted by $ P_{bolo} $, or the bolometric fluence, denoted by $ S_{bolo} $.
The isotropic luminosity is given by
\begin{equation}
 L_{\rm iso} = 4 \pi d^2_{L}(z) P_{\rm bolo} \,,
 \label{ldl}
\end{equation}
the total collimation-corrected energy reads as
\begin{equation}
E_{\rm iso}=4\pi d^2_{L}(z)S_{\rm bolo}(1+z)^{-1}\,,
\end{equation}
and the total collimation-corrected energy is
\begin{equation}
E_{\gamma}=4\pi d^2_{L}(z) S_{\rm bolo}F_{\rm beam}(1+z)^{-1}\,, \label{egdl}
\end{equation}
where $F_{\rm beam}$ is the beaming factor.
%%The correlation relations are power-law relations of either $ L_{\rm
%%iso}$ or $E_{\gamma}$ or $ E_{\rm iso}$ as a function of $\tau_{\rm lag}$, $V$,\, $E_{\rm peak}$,\, $\tau_{rt}$,
%%i.e.
%%\begin{eqnarray}\nonumber
%%&&E_{\rm iso} = b_{\rm iso,peak} E_{\rm peak}^{a_{\rm iso,peak}}\,,\nonumber\\
%%&&E_{\gamma} = b_{\gamma,peak}E_{\rm peak}^{a_{\gamma,peak}}\,,\nonumber\\
%%&&L = b_{peak}E_{\rm peak}^{a_{peak}}\,.\nonumber\\
%%\end{eqnarray}

Therefore, $L_{iso}$, $E_{\gamma}$ and $ E_{\rm iso}$ depend not only on the GRB observables $P_{\rm bolo}$ or
$S_{\rm bolo}$, but also on the cosmological parameters, through the luminosity distance $d_L(z)$. As a
consequence, there is a big circularity problem to overcome, since it is not immediately possible to calibrate such GRBs
empirical laws, and to build up a new GRBs Hubble diagram, without assuming any a priori cosmological model.
In  \cite{ME} we have applied a local
regression technique to estimate, in a model independent way, the distance modulus from the recently updated
Union SNIa sample, containing $557$ SNIa spanning the redshift range of $0.015 \le z \le 1.55$. The derived
calibration parameters have been used to construct an updated GRBs Hubble diagram. In particular, by using such
a technique, we have fitted the so-called \textit{Amati relation} and constructed an updated Gamma Ray Bursts
Hubble diagram, which we call the \textit{calibrated} GRBs HD, consisting of a sample of $109$ objects, shown in Fig. \ref{figrbs}. Their redshift distribution covers a broad range of $z$, from $0.033$ to $8.26$, thus extending far beyond that of SNIa ($z_{max} \sim 1.7$), and including GRB $092304$, the new high-z record holder of Gamma Ray Bursts.
\begin{figure}
\includegraphics[width=8 cm, height=6 cm]{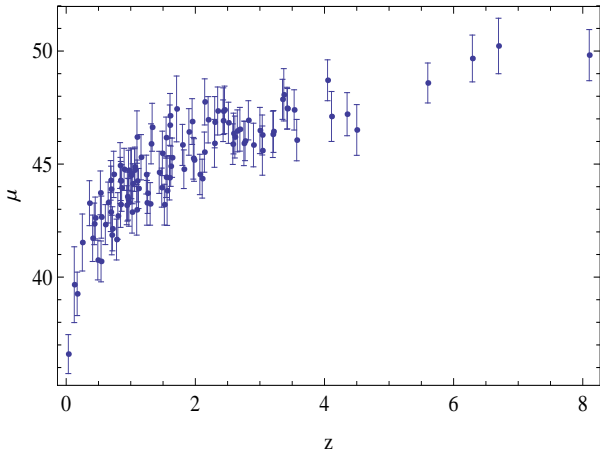}
\caption{Distance modulus $\mu(z)$ for the \textit{calibrated}  GRBs Hubble diagram made up by fitting the Amati
correlation.} \label{figrbs}
\end{figure}
Since here we want to use such \textit{calibrated} GRBs HD  to perform the high-redshift cosmographic analysis described below,
 it is worth to discuss some arguments about the reliability of using the Amati relation for cosmological tasks. For instance, one of the discussed objection is a supposition that the Amati relation is dominated by various selection effects of the detectors and differences within the GRB population. Recently such a thesis has been developed according to a conceptually simple argument based on the fact that the $S_{bolo}$-$E_{\rm peak}$ diagram presents two limit lines, where bursts cannot be below if the Amati and Ghirlanda relation holds \cite{n-p2005}, \cite{c-s2012}. Actually it turns out that
 \begin{equation}
\xi=\frac{E^{{\eta}^{-1}}_{peak}}{S_{bolo}} =  K^{\frac{1}{\eta}} 10^{\frac{\sigma}{\eta}}\frac{4 \pi d_{L}^2}{(1+z)^{1+\frac{1}{\eta}}},
\label{NaP}
\end{equation}
where $K$ and $\eta$ are the normalization and the slope of the Amati corralation ($E_{peak}= K E_{iso}^{\eta}$), and $\sigma$ corresponds to the scatter of the data points around the rest frame correlation. Let us note that the left side of the Eq. (\ref{NaP}) uses only directly observable quantities, while the right side is only a function of distance.  As the distance rises, $d_L^2$ gets larger and $(1+z)^{-(1+{\eta}^{-1})}$ gets smaller, giving rise to a maximum value for the right side, which cannot exceed therefore a limit value.
Indeed an immediate test of the Amati relation can be performed, by investigating where the average burst falls below the Amati limit.
In \cite{c-s2011}  the results of such an analysis have been interpreted as a clear disproof of the Amati relation. However  such a conclusion sounds premature, and should be postponed untill the available datasets will be more consistent  and once some  \textit{sources of uncertainty} have been taken under control. For instance, the \textit{limit lines } could vary with the background cosmology, or with the slope parameter, $\eta$, in a not negligible way, as indicated by Fig. \ref{energyratio}, and Fig. \ref{energy ratio2} with sensitive effects on the fraction of violators of such a limit ( on which the analysis itself is based).

  \begin{figure}
\includegraphics[width=6 cm, height=4 cm]{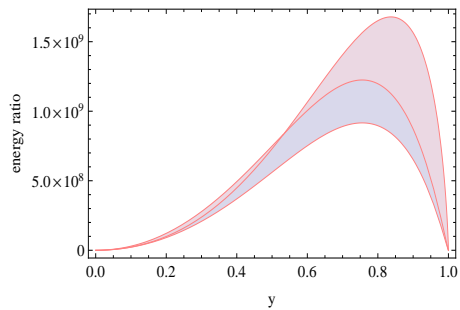}
\caption{Predicted value of the Amati relation energy ratio $\xi$ as a
function of redshift, varying the cosmological parameters and the slope of the $E_{iso}$-$E_{\rm peak}$. Here is illustrated an example of the $\Lambda$CDM model where the Hubble constant $h$ varies from $h=0.74$ to $h=0.65$ (narrow blue region), and the slope $\eta^{-1}$ from $\eta^{-1}=1.8$ to $\eta^{-1}=2.18$ (large pink region). It turns out that the peak can be shifted noticeably. } \label{energyratio}
\end{figure}

Here we do not intend to analyse thoroughly such aspects, which we postpone to a forthcoming paper, and which, even if important, do not touch the heart of our cosmographic investigation; however we want just to note, as an \textit{a posterior argument}, that if we relay on the Amati relation, and use it to build up the Hubble diagram, it turns out that it is fully consistent with the HD obtained from all the other relations available for the GRBs, as seen in Figs. \ref{amativalidity}. In Fig.  \ref{amativalidity2}, we show how such limit lines can be shifted, by varying the energy ratio as in the  Figure \ref{energyratio}. It turns out that the number of outliers strongly depends on the position of the limit lines. Moreover, in \cite{nava11} the spectral properties of short and long GRBs, detected by the {\it Gamma-ray Burst Monitor (GBM)}, have been studied over an unprecedented wide energy range. It turns out that the fraction of long GRBs, which are outliers (at more than $ 3 \sigma $) with respect to the Amati relation, is $ \sim 3 $ per cent, while there are no outliers (at more than $3 \sigma$) for the $E_{peak}-L_{iso}$ correlation.
\begin{figure}
\includegraphics[width=6 cm, height=4 cm]{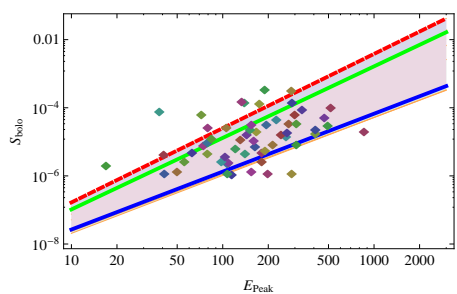}
\caption{The basics of the Nakar and Piran test in a graphical form. Any burst (even without a known redshift) can be plotted on this diagram. If the Amati  relation is correct, then all burst should lie above the solid line. Here it is shown how such a limit value can be shifted, by varying the energy ratio as in the previous Figure. The superposed points are taken from Wei 2010 ; it turns out that the number of outliers strongly depends on the position of the limit lines.} \label{energy ratio2}
\end{figure}

\begin{figure}
\includegraphics[width=10 cm, height=5 cm]{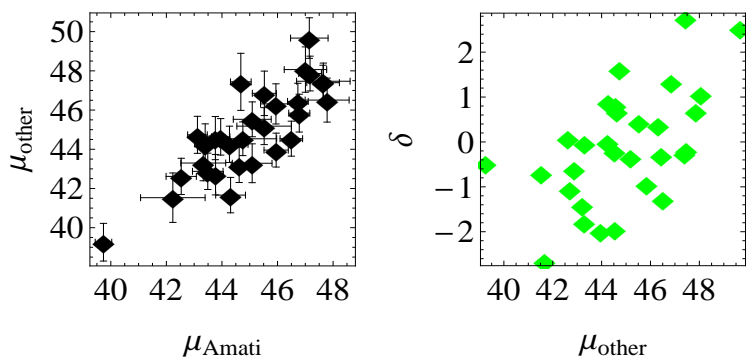}
\caption{{\bf Left panel}: comparison between the HD obtained with the Amati relation and the $L-\tau_{lag}$, $L-V$, $L-E_{peak}$, $L-\tau_{RT}$ relation. {\bf Right panel:} behaviour of the residuals among the Amati and the \textit{other} relations. It turns out that these datasets are fully consistent and strongly  correlated  with the Spearman�s correlation $\rho = 0.8$.} \label{amativalidity}
\end{figure}
 \begin{figure}
\includegraphics[width=10 cm, height=5 cm]{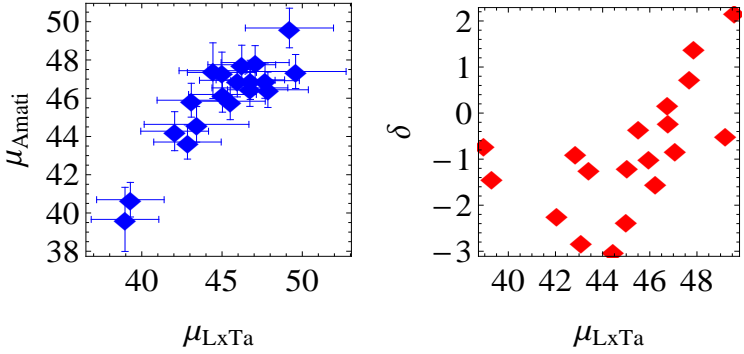}
\caption{{\bf Left panel}: comparison between the HD obtained with the Amati relation and the $L_X$\,-\,$T_a$ relation. {\bf Right panel:} behaviour of the residuals. It turns out that these data sets are fully consistent and strongly  correlated  with the Spearman�s correlation $\rho = 0.72$.} \label{amativalidity2}
\end{figure}
\subsubsection{The $L_X$\,-\,$T_a$  Gamma Ray Bursts Hubble diagram}
The $L_X$\,-\,$T_a$ correlation  between the luminosity $L_X$ at the break time $T_a$ and $T_a$ itself is the only empirical law relating quantities measured from the afterglow light curve, as described by the universal fitting function proposed by Willingale et al. (2007, hereafter W07), rather than being related to the prompt emission quantities. It has been first discovered by Dainotti et al. (2008) and later confirmed by the semiempirical models of Ghisellini et al. (2009) and Yamazaki (2009). More recently, Dainotti et al. 2010 ( herefater D10) have increased the GRBs sample and rederived the $L_X$\,-\,$T_a$ correlation, selecting a class of high luminosity long GRBs with very well measured $(L_X, T_a)$ parameters and lightcurve closely matching the W07 model. Referring to this class of objects as {\it canonical} GRBs, D10 have demonstrated that they define an upper envelope for the $L_X$\,-\,$T_a$ correlation with the same slope, but a higher intercept than that for the full sample. Cardone et al. (2011) used  this $L_X$\,-\,$T_a$  Gamma Ray Bursts Hubble diagram to constrain cosmological parameters of some simple dark energy models. Since the {\it canonical sample } is statistically  poor (it consists of only 8 objects), here we use the full sample to perform our cosmographic analysis. We \textit{build up} such sample calibrating, the $u < 4$ sample from D10, with Local Regression technique used in \cite{MEC10} and \cite{ME}, for the
\begin{equation}
\log{L_X} = a \log{\left ( \frac{T_a}{1 + z} \right )} + b
\label{eq:lxta}
\end{equation}
 correlation. Here $u$ is an error parameter: $u = \sqrt{\sigma_{L_X}^2 + \sigma_{T_a}^2}$.
In order to infer the distance modulus of each GRB, we then simply note that $L_X$ is related to the luminosity distance $d_L(z)$ as

\begin{equation}
L_X = 4 \pi d_L^2(z) (1 + z)^{-(2 + \beta)} F_X\,,
\label{eq:lxdl}
\end{equation}
$\beta$ being the slope of the energy spectrum (modelled as a simple power\,-\,law) and $F_X$ the observed flux both measured at the break time $T_a$. Having measured $(T_a, \beta, F_X)$ and inferred $L_X$ using Eq.(\ref{eq:lxta}), we can then estimate the GRB distance modulus as\,:

\begin{equation}
\mu(z) = 25 + 5 \log{d_L(z)} = 25 + \frac{5}{2} \log{\left [ \frac{L_X}{4 \pi (1 + z)^{-(2 + \beta)} F_X} \right ]}\,,
\end{equation}
where $d_L(z)$ is in Mpc. Such $(L_X, T_a)$-GRBs Hubble diagram in shown in Fig. \ref{cosmolxtahd}. The uncertainty is estimated by propagating the errors of $(\beta, F_X, L_X)$.
\begin{figure}
\includegraphics[width=6 cm, height=4 cm]{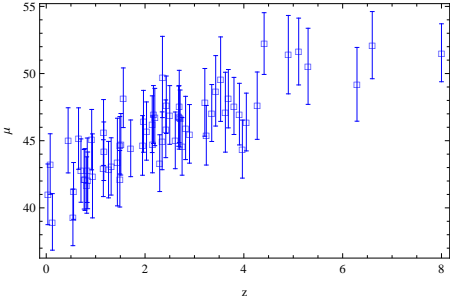}
\caption{Distance modulus $\mu(z)$ for the \textit{calibrated}  GRBs Hubble diagram made up by fitting the  $L_X-T_a$ correlation.} \label{cosmolxtahd}
\end{figure}
While both SNIa and GRBs are based on the concept of standard candles, an alternative way to probe the background evolution of the universe relies on the use of standard rulers.
Nowadays the Baryonic Acoustic Oscillations (BAOs) which are related to the imprint of the primordial acoustic waves on the galaxy power spectrum are widely used as such rulers.
In order to use BAOs as constraints, we follow \cite{P10} by first defining\,:
\begin{equation}
d_z = \frac{r_s(z_d)}{D_V(z)}\,,
\label{eq: defdz}
\end{equation}
with $z_d$ the drag redshift computed using the approximated formula in \cite{EH98}, $r_s(z)$ the comoving sound horizon given by\,:
\begin{equation}
r_s(z) = \frac{c}{\sqrt{3}} \int_{0}^{(1 + z)^{-1}}{\frac{da}{a^2 H(a) \sqrt{1 + (3/4) \Omega_b/\Omega_{\gamma}}}} \ ,
\label{defsoundhor}
\end{equation}
and $D_V(z)$ the volume distance defined by \cite{Eis05}\,:
\begin{equation}
D_V(z) = \left \{ \frac{c z}{H(z)} \left [ \frac{D_L(z)}{1 + z} \right ]^2 \right \}^{1/3} \ .
\label{defdv}
\end{equation}

\section{Statistical Analysis}

In this section we describe some details of our statistical analysis and present our main results on the constraints for the
cosmographic expansion parameters from the current observational data sets described above. In order to constrain the cosmographic parameters, we perform a preliminary and \textit{standard fitting} procedure to maximize the likelihood function ${\cal{L}}({\bf p}) \propto \exp{[-\chi^2({\bf p})/2]}$, where ${\bf p}$ is the set of cosmographic parameters and the expression for $\chi^2({\bf p})$ depends on the data set used.  As a first test we consider only the SNIa data, thus we define\,:
\begin{eqnarray}
\chi^2({\bf p}) & = & \sum_{i = 1}^{{\cal{N}}_{SNIa}}{\left [ \frac{\mu_{obs}(z_i) - \mu_{th}(z_i, {\bf p})}{\sigma_i} \right ]^2} \nonumber \\
& + &  \left ( \frac{h - 0.742}{0.036} \right )^2
+ \left ( \frac{\omega_m - 0.1356}{0.0034} \right )^2 \ .
\label{defchiSNIa}
\end{eqnarray}
Here, $\mu_{obs}$ and $\mu_{th}$ are the observed and theoretically predicted values of the distance modulus, while the sum is over all the SNIa  in the sample. The last two terms are Gaussian priors on $h$ and $\omega_M = \Omega_M h^2$ and are included in order to help break the degeneracies among the model parameters. To this aim, we have resorted to the results of the SHOES collaboration \cite{shoes} and the WMAP7 data \cite{WMAP7}, respectively, to set the numbers used in Eqs. (\ref{defchiSNIa}).
When we are using GRBs only, we define\,:
\begin{eqnarray}
\chi^2({\bf p}) & = & \sum_{i = 1}^{{\cal{N}}_{GRBHD}}{\left [ \frac{\mu_{obs}(z_i) - \mu_{th}(z_i, {\bf p})}{\sigma_i} \right ]^2} \nonumber \\
& + &  \left ( \frac{h - 0.742}{0.036} \right )^2
+ \left ( \frac{\omega_m - 0.1356}{0.0034} \right )^2 \ .
\label{chigrb}
\end{eqnarray}

As a next step, we combine the SNIa and GRBs HDs with other data redefining ${\cal{L}}({\bf p})$ as\,:
\begin{eqnarray}
{\cal{L}}({\bf p}) & \propto & \frac{\exp{(-\chi^2_{SNIa/GRB}/2)}}{(2 \pi)^{\frac{{\cal{N}}_{SNIa/GRB}}{2}} |{\bf C}_{SNIa/GRB}|^{1/2}} \nonumber \\
~ & \times  & \frac{1}{\sqrt{2 \pi \sigma_h^2}} \exp{\left [ - \frac{1}{2} \left ( \frac{h - h_{obs}}{\sigma_h} \right )^2
\right ]} \nonumber \\
~ & \times & \frac{\exp{(-\chi^2_{BAO}/2})}{(2 \pi)^{{\cal{N}}_{BAO}/2} |{\bf C}_{BAO}|^{1/2}} \nonumber \\
~ & \times & \frac{1}{\sqrt{2 \pi \sigma_{{\cal{R}}}^2}} \exp{\left [ - \frac{1}{2} \left ( \frac{{\cal{R}} - {\cal{R}}_{obs}}{\sigma_{{\cal{R}}}} \right )^2 \right ]} \nonumber \\
~ & \times & \frac{\exp{(-\chi^2_{H}/2})}{(2 \pi)^{{\cal{N}}_{H}/2} |{\bf C}_{H}|^{1/2}} \  .
\label{defchiall}
\end{eqnarray}

The first two terms are the same as above with ${\bf C}_{SNIa/GRB}$ the SNIa/GRBs diagonal covariance matrix and $(h_{obs}, \sigma_h) = (0.742, 0.036)$. The third term takes into account the constraints on $d_z = r_s(z_d)/D_V(z)$ with $r_s(z_d)$ the comoving sound horizon at the drag redshift $z_d$ (which we fix to be $r_s(z_d) = 152.6 \ {\rm Mpc}$ from WMAP7) and the volume distance is defined as in Eq. (\ref{defdv}).
The values of $d_z$ at $z = 0.20$ and $z = 0.35$ have been estimated by Percival et al. (2010) using the SDSS DR7 galaxy sample so that we define $\chi^2_{BAO} = {\bf D}^T {\bf C}_{BAO}^{-1} {\bf C}$ with ${\bf D}^T = (d_{0.2}^{obs} - d_{0.2}^{th}, d_{0.35}^{obs} - d_{0.35}^{th})$ and ${\bf C}_{BAO}$ is the BAO covariance matrix. The next term refers to the shift parameter \cite{B97,EB99}\,:

\begin{equation}
{\cal{R}} = H_{0} \sqrt{\Omega_M} \int_{0}^{z_{\star}}{\frac{dz'}{H(z')}}\,,
\label{eq: defshiftpar}
\end{equation}
with $z_\star = 1090.10$ the redshift of the last scattering surface. We follow again the WMAP7 data setting $({\cal{R}}_{obs}, \sigma_{{\cal{R}}}) = (1.725, 0.019)$. While all these quantities (except for the Gaussian prior on $h$) mainly involve the integrated $E(z)$, the last term refers to the actual measurements of $H(z)$ from the differential age of passively evolving elliptical galaxies. We then use the data collected by Stern et al. (2010) giving the values of the Hubble parameter for ${\cal{N}}_H = 11$ different points over the redshift range $0.10 \le z \le 1.75$ with a diagonal covariance matrix. We finally perform our cosmographic analysis, considering a whole data set containing both the SNIa Union data set and the \textit{calibrated} GBRs HD (which we call the \textit{cosmographic dataset}), and slightly modifying the likelihood ${\cal{L}}({\bf p})$.
To compute the likelihood, we use a  Bayesian approach, implementing a Monte Carlo Markov Chain technique. For each Monte Carlo Markov Chain calculation, we run two times four independent chains that consist  of about $100 000$ chain elements each \footnote{Actually multiple independent chains can be started in different points of the parameter space to ensure good mixing, i.e. an adequate exploration of the whole parameter space.}. We test the convergence of the chains by the Gelman and Rubin criterion, finding $R - 1$ of order $0.01$, which is more restrictive than the often used and recommended value $R - 1 < 0.1$ for standard cosmological investigations.
Moreover in order to reduce the uncertainties on cosmographic parameters, since methods like the MCMC are based on an algorithm that moves randomly in the parameter space, we \textit{a priori} imposed some constraints on the series expansions of $H^{2}(z)$ and $d_{L}(z)$.
The most general and obvious constraint is the positivity requirement\footnote{Let us note that here we do not use the constraint $0<\Omega_m<1$\,, that was employed, for instance, in \cite{salzcosmo}, since it is a sort of \textit{meta cosmographic} constraint, since it would require to \textit{postulate} the form of the Friedman equations. However, when in the following section we are going to investigate the implications of our results on the evolution of the dark energy equation of state for some parametrizations, we use such a constraint as an \textit{a posteriori} control.  }:
\begin{itemize}
 \item $d_{L}(z) > 0$ \,,
 \item $H^{2}(z) > 0$ \,,
\end{itemize}
applied for all our redshift ranges.
We first run our chains to compute the likelihood in Eqs. (\ref{defchiSNIa}) and/or (\ref{chigrb}), using as starting points the best fit values obtained  in our \textit{pre-statistical analysis}, in order to \textit{select} more efficiently the space (cosmographic) parameters region, and mainly to select the starting points. Therefore we perform the same Monte Carlo Markov Chain calculation to evaluate the likelihood in Eq. (\ref{defchiall}), combining the SNIa HD, the BAO and $H(z)$ data with the GRBs HD respectively, as described above. We will refer to such two different cases as Cosmography I and Cosmography II.
we throw away  first $30\%$  of the points iterations at
the beginning of any MCMC run, and we thin the two-runned chains.
We finally   extract the constraints on the parameters, coadding the thinned chains. Both Cosmography I and Cosmography II are implemented using $z$ (z-Cosmography I/II) and $y$ (y-Cosmography I/II) series. In Table \ref{tab1} we present the results of our cosmographic analysis applied to the \textit{Union2} SNIa data set. It turns out that in Cosmography I the snap parameter, $s_0$, is weakly constrained and the lerk, $l_0$, is actually unconstrained.
\begin{table}
\begin{center}
\begin{tabular}{c|c|c|c|c|c|c|}
  \hline
\hline
  Parameter&$h$&$q_0$&$j_0$&$s_0$&$l_0$\\
  \hline
  \hline
  Best Fit & $0.755$&$-0.462$ & $0.134$ & $0.696$ & $-3.57$\\
  Mean & $0.761$&$ -0.471$&$0.258$&$0.164$&$ -0.453$\\
  2 $\sigma$ & $ (0.71, 0.81)$&$ (-0.53, -0.40)$&$(0.105, 0.68)$&$(-0.996, 1.28)$&$(-8.23, 7.37)$\\
  \hline
  \hline
\end{tabular}
\end{center}
\caption{Constraints on the parameters of Cosmography I (from
combining the SNIa  HDs with BAO and $H(z)$  data set ($2\sigma$ error bars)).}
\label{tab1}
\end{table}
 It turns out that with the y-Cosmography I we obtain practically the same results, without any statistical meaningfull difference. However the $y$-redshift approach allows a faster convergence of the chains.

In the same way, the z-Cosmography II and y-Cosmography II turned out to be fully statistically equivalent and in Table \ref{tab2} we present the results of our cosmographic analysis adding the \textit{calibrated Amati} Gamma Ray Bursts Hubble diagram in the $y$-redshift case (y-Cosmography II). In Fig. \ref{like2} we plot the marginalized likelihood function for the deceleration parameter $q_0$. It turns out that in this  Cosmography II the snap parameter, $s_0$, is weakly constrained and the lerk, $l_0$, is actually unconstrained.

\begin{table}
\begin{center}
\begin{tabular}{c|c|c|c|c|c|c|}
  \hline
\hline
  Parameter&$h$&$q_0$&$j_0$&$s_0$&$l_0$\\
  \hline
  \hline
  Best Fit&$0.742$&$-0.406$&$ 0.86$&$29.767$&$-7.757$\\
  Mean&$0.732$&$-0.359$&$ 0.28$&$20.33$&$1.044$\\
 2 $\sigma$ &$(0.67, 0.81)$&$(-0.53, -0.134)$&$(0.104, 0.95)$&$(-0.86, 29.72)$&$ (-30.01, 29.95)$\\
  \hline
  \hline
\end{tabular}
\end{center}
\caption{Constraints on the parameters of the Cosmography II (from
combining the SNIa HD, the \textit{ Amati} Gamma Ray Bursts HD with BAO and $H(z)$  data sets ($2\sigma$ error bars)).}
\label{tab2}
\end{table}

\begin{figure}
\includegraphics[width=6 cm, height=4 cm]{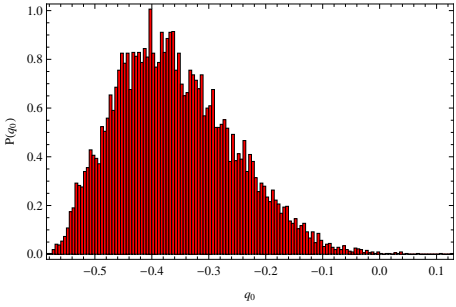}
\caption{The marginalized probability density function (PDF) for the deceleration parameter $q_0$, as provided by  Cosmography II.} \label{like2}
\end{figure}

It is worth noting that from our statistical MCMC analysis it turns out that the deceleration parameters $q_0$ is clearly negative in all the cases. The marginal likelihood distribution for the current deceleration parameter $q_0$ indicates that there is just a negligible probability for $q_0>0$, as shown in Fig. \ref{like2}. Moreover the value of the jerk $j_0$ is significantly different from the $\Lambda$CDM value $j_0=1$, what is also indicated by the marginal likelihood distribution, see Fig. \ref{likej0}.
\begin{figure}
\includegraphics[width=8 cm, height=5 cm]{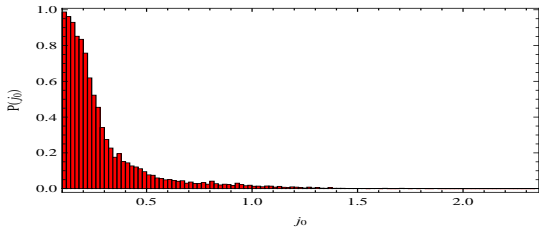}
\caption{The marginalized probability density function (PDF) for the jerk parameter $j_0$, as provided by  Cosmography II.} \label{likej0}
\end{figure}
In Fig. \ref{conregq0j0} are shown the confidence regions for $h$, $q_0$, and $j_0$: the left-side and the right-side panels concern the ($h$-$q_0$) and the ($q_0$-$j_0$) plane, respectively.
\begin{figure}
\includegraphics[width=8 cm, height=4cm]{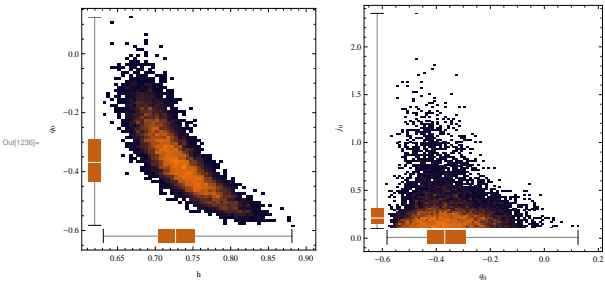}
\caption{Confidence regions in the ($h$-$q_0$) and the ($q_0$-$j_0$) plane, respectively, as provided by  Cosmography II. The inner brown region define the 3\,$\sigma$ confidence level. It turns out that the parameters $h$, $q_0$ and $j_0$ are well-constrained, that the values $q_0>0$ are ruled out, the value $j_0=1$ (which is the  $\Lambda$CDM value) is statistically not favourable.} \label{conregq0j0}
\end{figure}
In Fig. \ref{fit} we plot the observational data compared with the maximum likelihood curve.
\begin{figure}
\includegraphics[width=6 cm, height=4 cm]{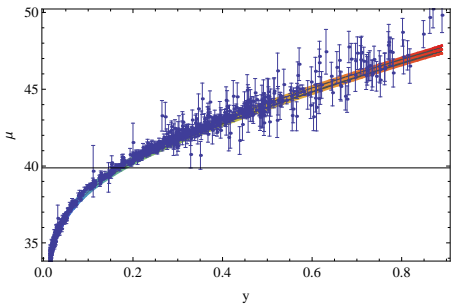}
\caption{Comparison of the observational data used in our Cosmography II  with the maximum likelihood curve (the thick red curve) visible in the center of the filled region, corresponding to the $2 \sigma$ confidence levels for the cosmographic parameters.} \label{fit}
\end{figure}

Independently of the cosmic deceleration today, it is of interest to investigate
if there is another change in the sign of the cosmic acceleration after a
prior transition from a decelerated to an accelerated phase at moderate
redshifts ($z_t\sim 0.5-1$). We use the cosmographic parameters ($q_0$, $j_0$, $s_0$ and $l_0$) to reconstruct $q(y)$ and to trace the deceleration history of the universe. Actually, from the power series expansion of the scale factor one can
also express the deceleration parameter as a power series in time, according to the definition in Eq. \ref{eq:cosmopar}.
This time dependent parameter can be written as a power series in
{\it y-redshift}, using the derivation rule
\begin{equation}\label{dty}
\frac{d}{dt} = (1 - y) H \frac{d}{dy}\,.
\end{equation}
The derivation of a power expansion (of fourth order) for $q(y)$ from
the scale factor expansion allows for a decelerated past, a transition to an accelerated phase,
a point of maximum acceleration, then a slowing down of the acceleration and
a transition to a recent or future decelerating phase, as shown in Fig. \ref{qy}. Appearance  of transient acceleration is predicted or allowed by several
dynamic models \cite{2009PhRvD}.
In contrast, the $\Lambda$CDM model predicts a monotonic deceleration history
connecting its asymptotic limits in the past and future,
$q(z\rightarrow \infty)=0.5$ and $q(z\rightarrow -1)=-1$.
The $q(z)$ reconstruction obtained from our cosmographic parametrization is
shown in figure \ref{qy}.
\begin{figure}
\includegraphics[width=6 cm, height=4 cm]{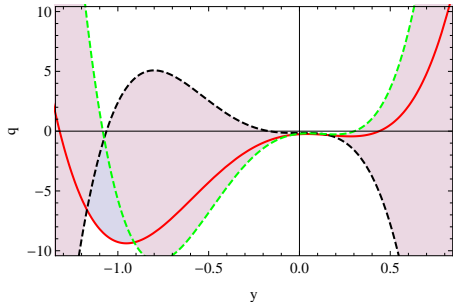}
\caption{Reconstruction of deceleration history.
The filled area delimits the 2$\sigma$
confidence region for the $q(z)$ reconstruction obtained from our Cosmography II.} \label{qy}
\end{figure}
It is worth noting that also the $q(z)$ reconstruction obtained from our Cosmography I allows a transient acceleration, as shown in Fig. \ref{qysneia}, but the specific properties of the deceleration history of the universe is rather different from the one obtained if we include GRBs HD in our cosmographic parametrization.
\begin{figure}
\includegraphics[width=6 cm, height=4 cm]{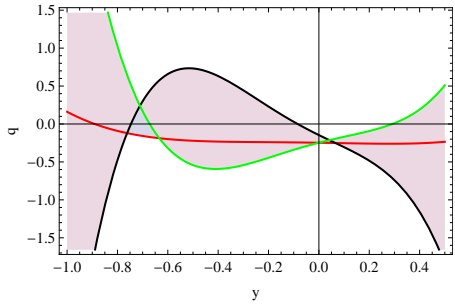}
\caption{Reconstruction of deceleration history obtained from our Cosmography I.} \label{qysneia}
\end{figure}

\subsection{Cosmography with the $L_X$\,-\,$T_a$  Gamma Ray Bursts Hubble diagram}
In this section we shortly
describe the results of our statistical analysis of another data set formed by combining the BAO and $H(z)$ data with the  $L_X$\,-\,$T_a$  Gamma Ray Bursts Hubble diagram described above (referred in the following as Cosmography III). We have decided not to include such GRBs data set in the overall  analysis, performed in Cosmography II, by virtue of the uncertain features of this recently discovered correlation, which should be further investigated in order to get stronger confidence and confirmation. It turns out that the results are mostly compatible with the previous ones obtained in our Cosmography I and II, as indicated in Tab. \ref{tab3}. Moreover it is worth noting that also in this case the value $j_0=1$ (which is the  $\Lambda$CDM value for the jerk) is statistically not favourable, as shown also in Fig. \ref{conregq0j0lxta}.

\begin{table}
\begin{center}
\begin{tabular}{c|c|c|c|c|c|c|}
  \hline
\hline
  Parameter&$h$&$q_0$&$j_0$&$s_0$&$l_0$\\
  \hline
  \hline
  Best Fit&$0.74$&$-0.44$&$ 0.386$&$-0.719$&$3.82$\\
  Mean&$0.713$&$ -0.402$&$ 0.45$&$2.54$&$-24.1$\\
 2 $\sigma$ &$(0.68, 0.79)$&$(-0.50, -0.26)$&$(-0.73, 1.4)$&$(-1.3, 8.4)$&$ (-72.0, 8.5)$\\
  \hline
  \hline
\end{tabular}
\end{center}
\caption{Constraints on the parameters of the Cosmography III (from
combining the SNIa HD, the $L_X$\,-\,$T_a$  Gamma Ray Bursts HD with BAO and $H(z)$  data sets ($2\sigma$ error bars)).}
\label{tab3}
\end{table}

\begin{figure}
\includegraphics[width=8 cm, height=4cm]{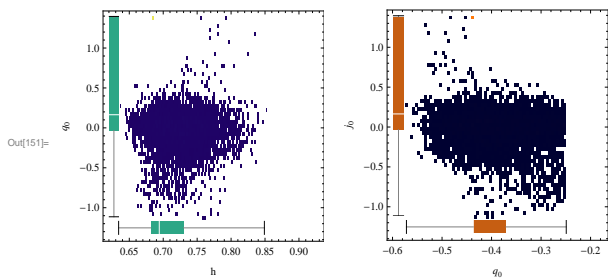}
\caption{Confidence regions in the ($h$-$q_0$) and the ($q_0$-$j_0$) plane, respectively, as provided by  the Cosmography III. It turns out that these results are compatible with the previous ones obtained in our Cosmography II.} \label{conregq0j0lxta}
\end{figure}
The $q(z)$ reconstruction allows a transient acceleration, shown in Fig. \ref{qylxta}, as in Cosmography I and II, strengthening the \textit{reliability } of the  $L_X$\,-\,$T_a$ correlation.
\begin{figure}
\includegraphics[width=6 cm, height=4 cm]{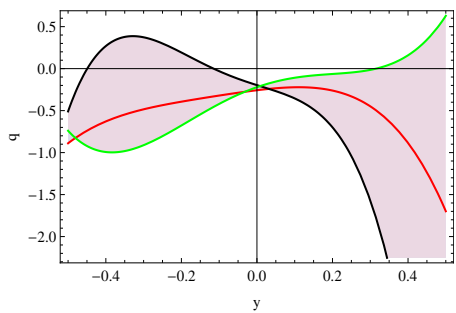}
\caption{Reconstruction of deceleration history obtained from our Cosmography III. The filled area delimits the 2$\sigma$
confidence region for the $q(z)$ reconstruction obtained from our analysis. } \label{qylxta}
\end{figure}
\section{Implications for Dark Energy}
In this section we investigate the implications of the results from our simulations on cosmography for different and specific parametrized dark energy models.
The link between the cosmographic and the dark energy parametrization is based on the series expansion (in redshift) of the Hubble function $H(z)$. Actually, for a spatially flat cosmological model
it turns out that:
\begin{eqnarray}
  H(z) &=& H_0 \sqrt{(1-\Omega_m) g(z)+\Omega_m (z+1)^3}\,,\\
 H_d(z) &=&- (z+1) H(z) H'(z)\,, \\
  H_{2 d}(z) &=& -(1+z) H(z) H_d'(z)\,, \\
  H_{3d}(z) &=& -(1+z) H(z)  H_{ 2 d}'(z)\,,\\
  H_{4d}(z) &=& -(1+z) H(z)  H_{ 3 d}'(z)\,,
\end{eqnarray}
 where $g(z)=\exp^{3 \int_0^z \frac{w(x)+1}{x+1} \, dx}$, and $w(z)$ any parametrized form of the dark energy equation of state.
It turns out that
\begin{eqnarray}
  \lim_{z->0}H_d(z) &=& -H_0 \left(1+q_0 \right)\label{deparcosmo1}\,, \\
    \lim_{z->0}H_{2 d}(z) &=& H_0^3 \left(j_0 + 3 q_0 + 2\right)\label{deparcosmo2}\,, \\
    \lim_{z->0}H_{3 d}(z)&=& H_0^4 \left(s_0 - 4 j_0 - 3 q_0 (q_0 + 4) - 6\right)\label{deparcosmo3}\,, \\
  \lim_{z->0} H_{4 d}(z) &=& H_0^5 \left(l_0 - 5 s_0 + 10 (q_0 + 2) j_0 + 30 (q_0 + 2) q_0 + 24\right)\label{deparcosmo4}\,.
\end{eqnarray}
 In this section we consider three different parametrizations:
\begin{itemize}
 \item the CPL parametrization for dark energy given by
\begin{equation}
w(z) =w_0 + w_{1} z (1 + z)^{-1} \,,
\label{cpleos}
\end{equation}
\item  a novel parametrization recently introduced in \cite{novel} to avoid the future divergency
problem of the CPL parametrization, and to probe the
dynamics of dark energy not only in the past evolution but also
in the future evolution,
\begin{equation}\label{noveleos}
w(z)=w_0+w_1 \left(\frac{\sin (z+1)}{z+1}-\sin (1)\right)\,,
\end{equation}

\item an oscillating dark energy equation of state recently discussed in \cite{salzoscil}
\begin{equation}
 w(z)=\frac{w_1 (1-\cos (\delta  \log (z+1)))}{\log (z+1)}+w_0\,.\label{osceos}
\end{equation}
\end{itemize}
It is worth noting that the efficiency of such investigation, i.e the possibility of inverting the equations  (\ref{deparcosmo1}--\ref{deparcosmo4}), strongly depends on the number of cosmographic parameters we are working with and on how many DE equation of state parameters we
are going to consider as free. For instance, restricting to the CPL equation of state given in Eq. \ref{cpleos}, one has three possibilities:
\begin{itemize}
\item with two cosmographic parameters, $(q_{0}, j_{0})$, we can
derive some information about a constant dark energy model (i.e. $w_{1} =
0$), with:
{\setlength\arraycolsep{0.2pt}
\begin{eqnarray}
\Omega_{m}(q_{0}, j_{0}) &=& \frac{2 (j_{0} - q_{0} - 2 q_{0}^{2})}{1 + 2 j_{0}
- 6 q_{0}}\, , \nonumber \\
w_{0}(q_{0}, j_{0}) &=& \frac{1 + 2 j_{0} - 6 q_{0}}{-3 + 6 q_{0}}\,,
\end{eqnarray}}
\item with two cosmographic parameters, $(q_{0}, j_{0})$, we can
derive some information also about a dynamical dark energy model, (i.e. $w_{1} \neq
0$), leaving $\Omega_{m}$ free, with:
{\setlength\arraycolsep{0.2pt}
\begin{eqnarray}
w_{0}(q_{0},
\Omega_{m}) &=& \frac{1 - 2 q_{0}}{3 (-1 + \Omega_{m})}\, , \nonumber \\
w_{1}(q_{0}, j_{0}, \Omega_{m}) &=& \frac{1}{3 (-1 + \Omega_{m})^{2}} \left(-2j_{0}(-1+\Omega_{m})-2q_{0}
\times  (1+2q_{0}) + \Omega_{m}(-1+6q_{0})\right)\,,
\end{eqnarray}}
\item with three cosmographic parameters, $(q_{0}, j_{0}, s_{0})$,
we can derive some information about a dynamical dark energy model,
with $\Omega_{m}$ depending on the cosmographic parameters, i.e.
$\Omega_{m} \doteq \Omega_{m}(q_{0}, j_{0}, s_{0})$. The same holds true for
$w_{0} \doteq w_{0}(q_{0}, j_{0}, s_{0})$ and $w_{1} \doteq w_{1}(q_{0}, j_{0}, s_{0})$.
\end{itemize}
For these relations all the statistical properties of these parameters (median, error bars, etc.) can
be directly extracted from the cosmographic samples we have obtained
from the MCMC analysis.
In our investigation we prefer to adopt a conservative approach, considering only the parameters which are well constrained by our cosmographic analysis, that is $(q_{0}$, and $ j_{0})$ leaving $\Omega_{m}$ free. It is therefore possible to obtain $w_0$ and $w_1$ in terms of $\Omega_m$ and other cosmographic parameters. For the parametrizations of the equation of states examined above, we get
\begin{itemize}
\item parametrization given by Eq. (\ref{noveleos})
\begin{eqnarray}
w_0(q_{0},\Omega_m) & = & \frac{1-2 q_0}{3 (\Omega_m-1)}\,,\\
w_1(q_{0}, j_{0},\Omega_m)& = & -\frac{2 j_0 \left(\Omega _m-1\right)+2 q_0 \left(-3 \Omega _m+2 q_0+1\right)+\Omega _m}{3 \left(\Omega _m-1\right){}^2 (\cos (1)-\sin (1))}\,,
\end{eqnarray}
\item parametrization given by Eq. (\ref{osceos})
\begin{eqnarray}
  w_0(q_{0},\Omega_m) &=& \frac{1-2 q_0}{3 \left(\Omega _m-1\right)}\\
  w_1(q_{0}, j_{0},\Omega_m) &=& \frac{-2 j_0 \left(\Omega _m-1\right)+\left(6 q_0-1\right) \Omega _m-2 q_0 \left(2 q_0+1\right)}{3 \delta ^2 \left(\Omega _m-1\right){}^2}\,.\label{w1osc}
\end{eqnarray}
\end{itemize}
In  Eq. \ref{w1osc} the parameter $\delta$ can be expressed as a function of $\Omega_m$ and $l_0$, inverting Eq. \ref{deparcosmo4}, which in our case gives
\begin{eqnarray}
 && l_0 = \frac{1}{4} \left(3 \left(\delta ^4 w_1 \left(\Omega _m-1\right) \left(3 w_1 \left(\Omega _m-7\right)+2\right)+\delta ^2 w_1 \left(9 w_0 \left(\Omega _m-1\right) \left(w_0
   \left(11 \Omega _m-23\right)+8 \Omega _m-31\right)-92 \Omega _m\right)\right.\right.+\nonumber\\
&& \left. \left. w_0 \left(\Omega _m-1\right) \left(3 w_0 \left(6 \left(6 w_0 \left(w_0+2\right)+7\right) \Omega _m-3
   w_0 \left(18 w_0+47\right)-134\right)-163\right)\right)+276 \delta ^2 w_1+70\right).
\end{eqnarray}

\begin{figure}
\includegraphics[width=9 cm, height=10cm]{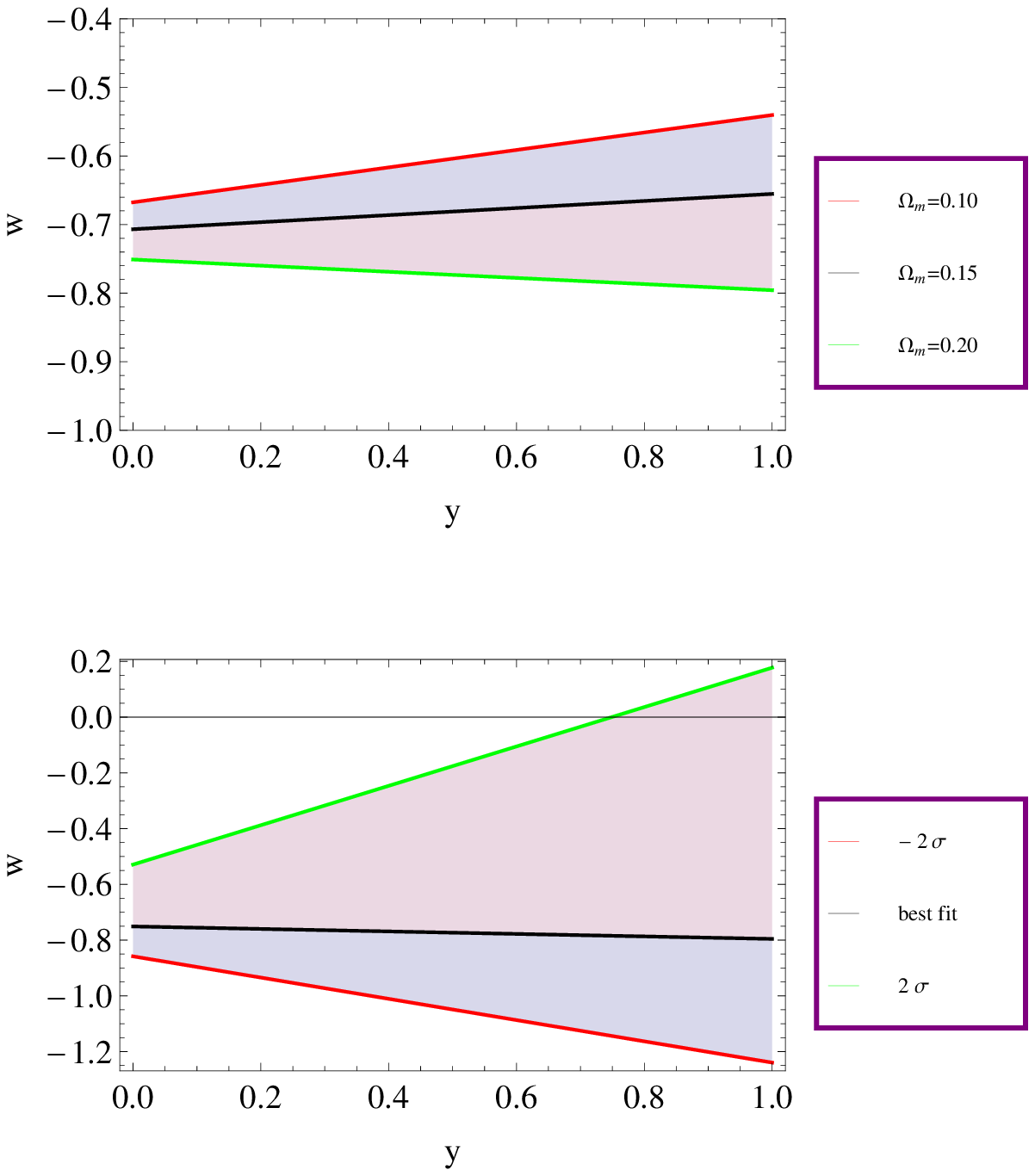}
\caption{{\bf  Upper Panel}
The $y$-redshift dependence of the equation of state described by the CPL parametrization (see Eq. \ref{cpleos}) for different values of
the $\Omega_m$, as described in the label.
{\bf Bottom Panel}
The $y$-redshift dependence of the CPL equation of state for different values of the cosmographic parameters $q_0$ and $j_0$.
The filled region corresponds to the \textit{allowed} behaviour of
the equation of state, when the cosmographic parameters are varying within the $2 \sigma$ region of confidence. $\Omega_m $ is fixed and
set to be $\Omega_m = 0.22 $.}\label{figcpleos3b}
\end{figure}

\begin{figure}
\includegraphics[width=9 cm, height=10cm]{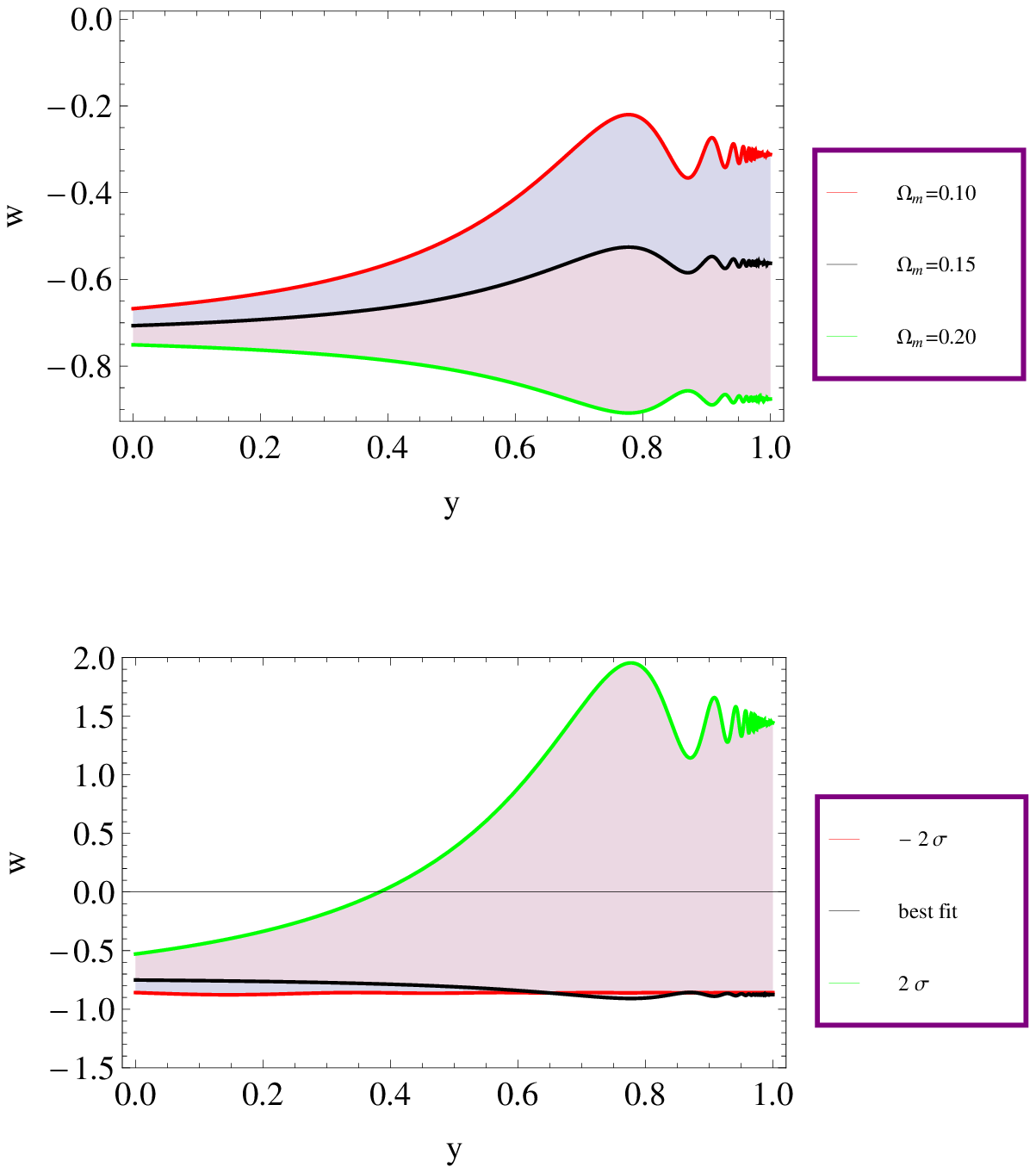}
\caption{{\bf  Upper Panel} The $y$-redshift dependence of the equation of state described by the parametrization (see Eq. \ref{noveleos}) for different values of the $\Omega_m$, as described in the label.
{\bf Bottom Panel}
The $y$-redshift dependence of the same equation of state for different values of the cosmographic parameters $q_0$ and $j_0$.
The filled region corresponds to the \textit{allowed} behaviour of
the equation of state, when the cosmographic parameters are varying within the $2 \sigma$ region of confidence. $\Omega_m $ is fixed and
set to be $\Omega_m = 0.22 $.} \label{fignoveleos3}
\end{figure}

\begin{figure}
\includegraphics[width=9 cm, height=10cm]{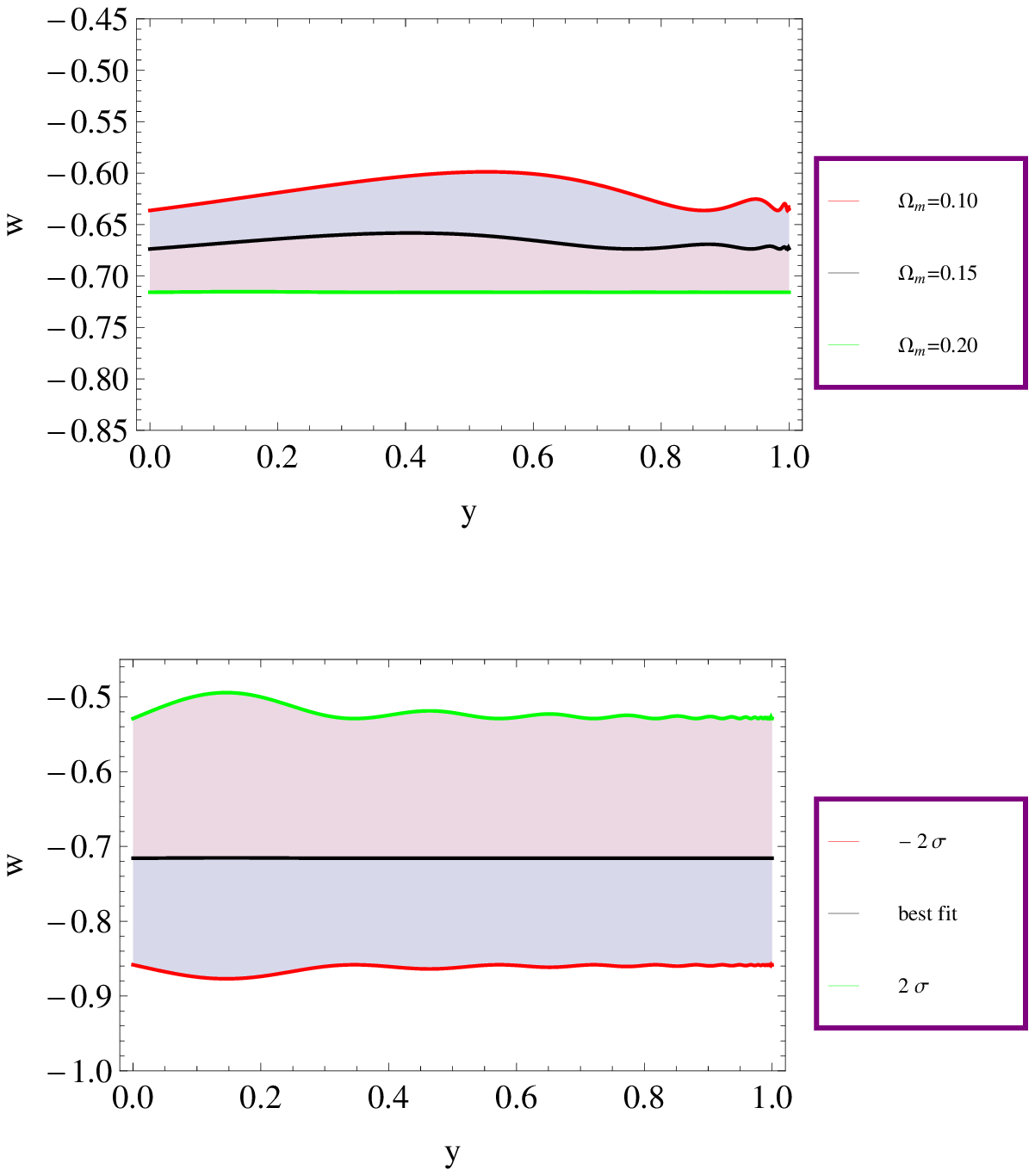}
\caption{{\bf  Upper Panel} Dependence of $w$ on the $y$-redshift in the equation of state described by the parametrization (see Eq. \ref{osceos}) for different values of the $\Omega_m$, as described in the label. {\bf Bottom Panel}
Dependence of $w$ on the $y$-redshift in the same equation of state for different values of the cosmographic parameters $q_0$ and $j_0$.
The filled region corresponds to the \textit{allowed} behaviour for
the equation of state, when the cosmographic parameters are varying within the $2 \sigma$ region of confidence. $\Omega_m $ is fixed and
set to the value $\Omega_m = 0.22 $.}
\label{figoscieos3}
\end{figure}
Actually, in Figs. \ref{figcpleos3b}, \ref{fignoveleos3}, and \ref{figoscieos3}
we can reconstruct the redshift behaviour of the different dark energy equations of state corresponding to different values of $\Omega_{m}$ (left panel), and obtained when the cosmographic parameters are varying within the $2 \sigma$ region of confidence,
and $\Omega_m$ is fixed (and set to be $\Omega_m= 0.22$).
The cosmographic analysis allows us to infer the actual value of the parameters appearing in the dark energy equation of state, and, as expected, can provide constraints on its redshift evolution mainly in a low redshift range. It turns out that the equation of state is evolving for all the parametrization considered, as confirmed for example in Figs. \ref{w1cplcontour}, where it is clear that for the parameter $w_1$, appearing in the \textit{non constant} term of the equation of state, the case $w_1=0$  has marginal confidence in the $2 \sigma$ region of confidence for the parameters $q_0,j_0$,  independently of the value of $\Omega_m$. Moreover, Fig. \ref{w1novelcontour}  reflects the possibility of a deviation from the $\Lambda$CDM cosmological model. 
\begin{figure},
\includegraphics[width=9 cm, height=10cm]{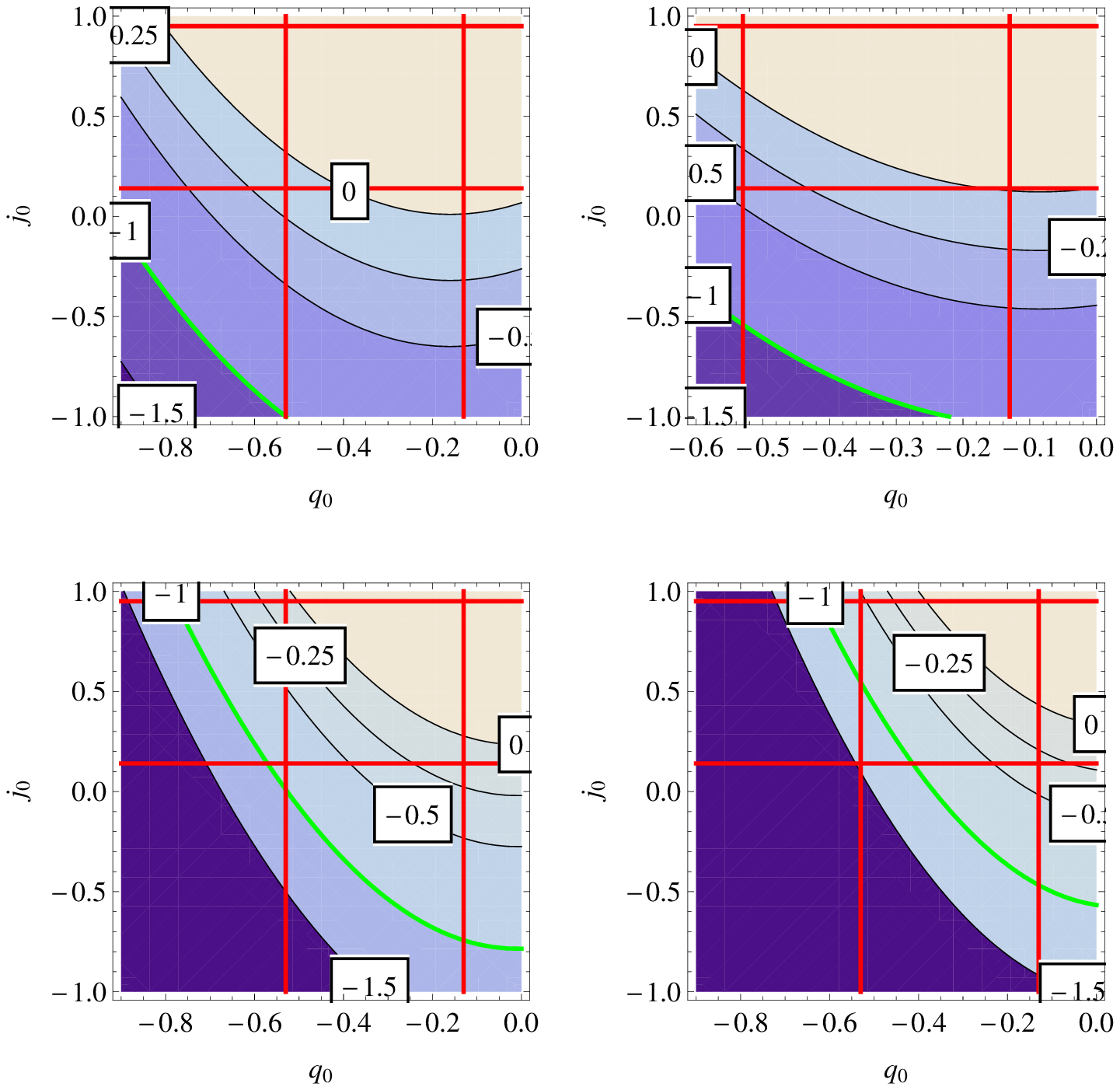}
\caption{Contour plots for the parameter $w_1(q_0,j_0)$, and for different and fixed values of $\Omega_m$, in the case of the CPL parametrization. The plots correspond to the values  $\Omega_m = 0.4$, $\Omega_m = 0.32$ (left and right side-upper panel), $\Omega_m=0.22$  and $\Omega_m=0.12$ (left and right side-bottom panel). The red lines define the  $2 \sigma$ region of confidence for the parameters $q_0,j_0$, as constrained by Cosmography II.  }
\label{w1cplcontour}
\end{figure}
Results of our cosmographic analysis are only marginally compatible with predictions of the $\Lambda$CDM model. In a forthcoming paper we are going to reconstruct the EOS of dark energy from the observational data.
\begin{figure}
\includegraphics[width=6 cm, height=5cm]{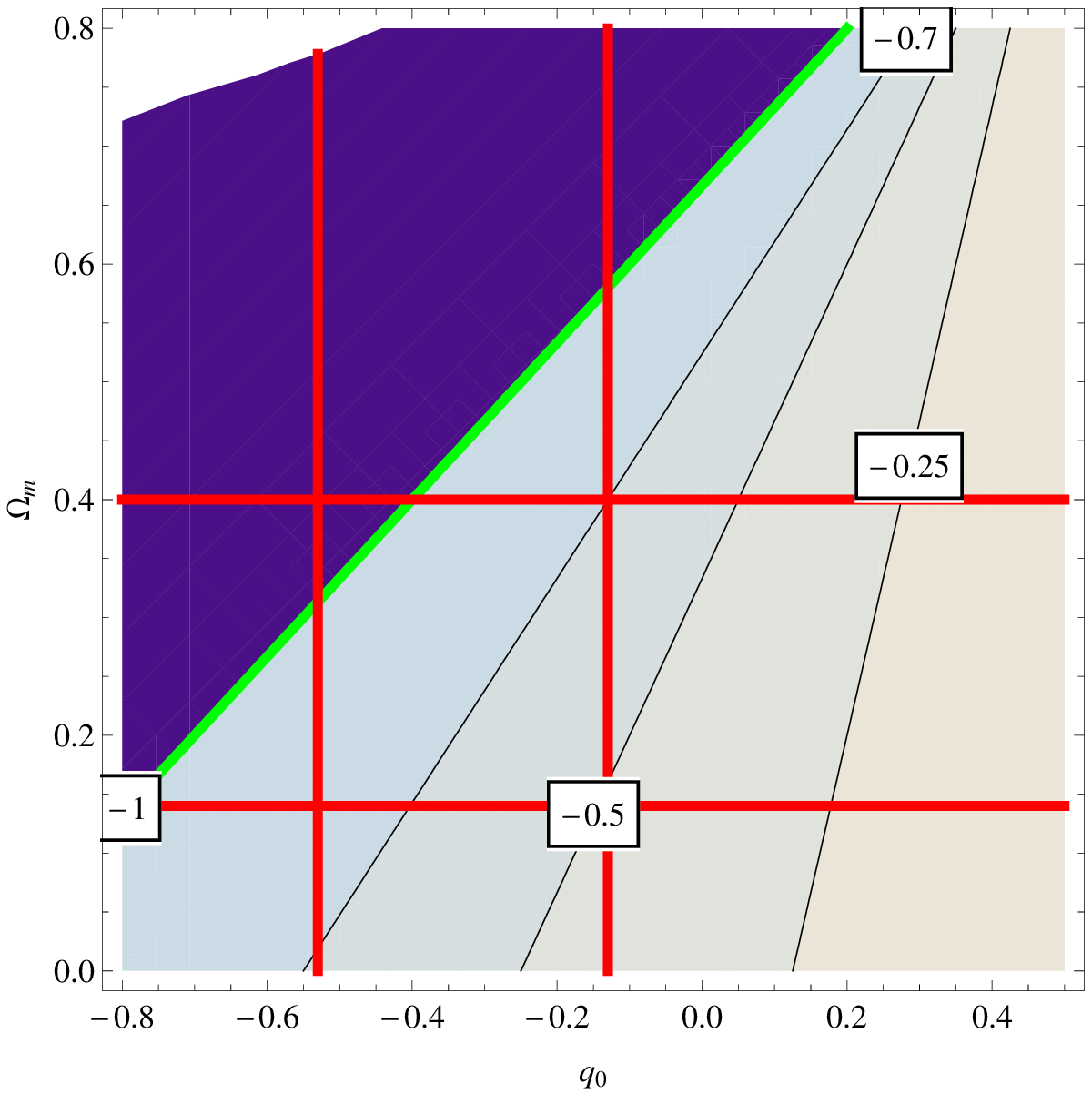}
\caption{Contour plot for the parameter $w_0(q_0,\Omega_m)$  in the case of the CPL parametrization. The red lines define the  $2 \sigma$ region of confidence for the parameters $q_0,\Omega_m$, as constrained by Cosmography II.  }
\label{w1novelcontour}
\end{figure}
As a final remark we note that the reliability of our cosmographic analysis
is strongly related to the question whether there is a relation between the
highest expansion order in the Taylor series and the redshift
range where this series can be applied. One could expect
that we would need a series expansion truncated at
higher orders when increasing up the redshift range.
However, the errors on the cosmographic coefficients will increase
when higher order expansions are considered. As those parameters are correlated among them,
errors in the low order series coefficients propagate to the
additional coefficients included in the higher order series. If
these errors turn out to be too large, the strength of cosmography will vanish.
In order to estimate the error resulting from stopping the expansion
at the fifth order we analyze the (relative) residuals between the exact distance modulus
and fifth order series expansions of the same
quantity, when varying the cosmographic parameters $q_0$, $j_0$, $s_0$, and $l_0$  within the $3 \sigma $ region of confidence, and for the CPL parametrization considered above, as illustrated in Fig. \ref{errortruncation}. It turns out that actually the error can be significant (with respect to the data)
already for $y\geq 0.8$ (corresponding to $z \simeq 4$). As far as  the results of our analysis are concerned, they are not dramatically affected by such \textit{truncation} error, not only because few data-points falls into the \textit{forbidden} region, but mainly since we expect to estimate  at most $q_0$, $j_0$ and $s_0$  (we actually consider $l_0$ essentially unbounded), that is the cosmography parameters connected to the expansion at forth order. Moreover when we investigate the implications of the results from our simulations on cosmography for different and specific parametrized dark energy model, we have considered only the parameters which are well constrained by our cosmographic analysis, that is $q_{0}$, and $ j_{0}$. Finally it is worth noting that Fig. \ref{errortruncation} suggests that, because of the actual precision of the observations, cosmography should not be used to calibrate the GRBs correlation relation, even if one limits the procedure to GRBs at $y<0.6$.
\begin{figure}
\includegraphics[width=8cm, height=4cm]{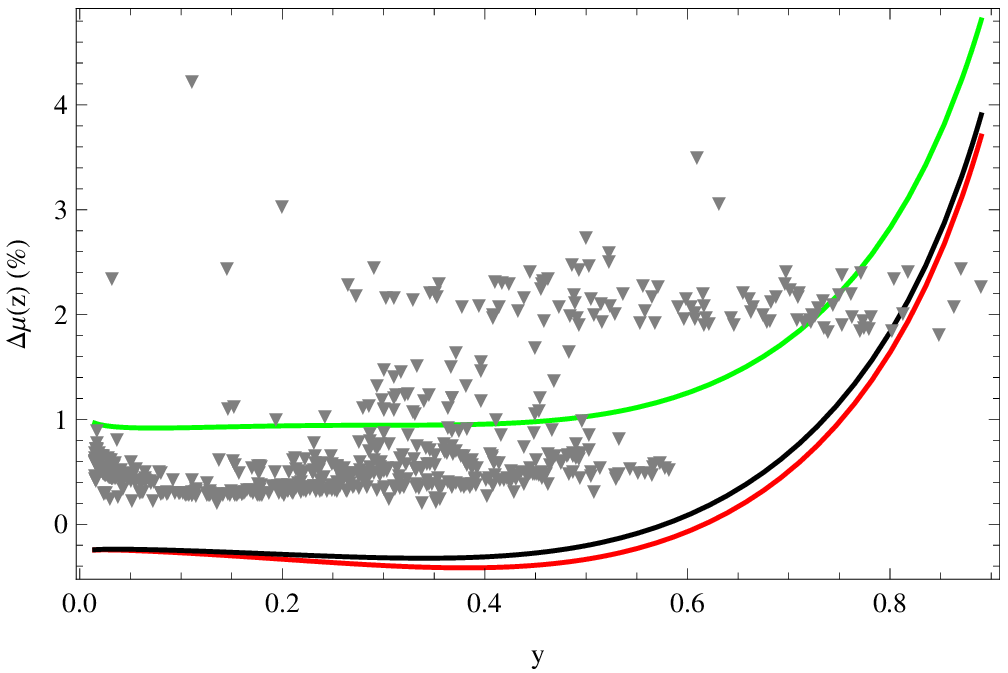}
\caption{Relative residuals between the CPL exactdistance modulus
and fifth order series expansions of the same quantity, when varying the cosmographic parameters $q_0$, $j_0$, $s_0$, and $l_0$ for the CPL parametrization considered above. The black thick line corresponds to the best fit values of the cosmographic values, while the red and green lines define  the $3 \sigma $ region of confidence. The grey triangles indicate the \textit{observational relative errors} for the SNeIa and GRBs data.  }
\label{errortruncation}
\end{figure}

\section{Discussion and Conclusions}
In this paper we are studying  the possibility to extract model independent information
about the dynamics of the universe by using a cosmographic approach considering only minimal assumptions (isotropy, homogeneity, Taylor series expansion of distances) without choosing any dynamical model a priori. In order to explore it systematically, we performed an high-redshift analysis that allowed us to put constraints on the cosmographic expansion up to the fifth order, based on the Union2 Type Ia Supernovae (SNIa) data set, the Hubble diagram constructed from some Gamma Ray Bursts luminosity distance indicators, and gaussian priors on the distance from the Baryon Acoustic Oscillations (BAO), and the Hubble constant $h$. Actually we use two GRB HD data set: one sample consists of $109$ high redshift GRBs and has been constructed from the
Amati $E_{\rm p,i}$ -- $E_{\rm iso}$ correlation. The second GRBs HD sample is constructed from 66 Gamma Ray Bursts (GRBs) derived using only data from their X\,-\,ray afterglow light curve. To this end, we used the recently updated $L_X$\,-\,$T_a$ correlation between the break time $T_a$ and the X\,-\,ray luminosity $L_X$ measured at $T_a$ calibrated (using SNIa) from a sample of {\it Swift} GRBs.
To reduce the uncertainties on cosmographic parameters, since methods like the MCMC are based on an algorithm that moves randomly in the parameter space, we \textit{a priori} imposed some constraints on the series expansions of $H^{2}(z)$ and $d_{L}(z)$, requiring that
the most general and obvious constraint is the positivity requirement:
\begin{itemize}
 \item $d_{L}(z) > 0$ \,,
 \item $H^{2}(z) > 0$ \,,
\end{itemize}
applied on all our redshift ranges.
We perform the same Monte Carlo Markov Chain calculations to evaluate the likelihoods, firstly considering the SNIa HD, the BAO and $H(z)$ data sets, or the GRBs-Amati HD, the BAO and $H(z)$ datasets separately (Cosmography I), and then constructing an overall data set joining them together (Cosmography II). Instead, we have decided not to include the  $L_X$\,-\,$T_a$  Gamma Ray Bursts Hubble diagram in the overall  analysis, performed in Cosmography II, by virtue of the peculiar features of this recently discovered correlation, which should be further investigated in order to get stronger confidence and confirmation.
Either one of the cosmographic analysis are implemented using $z$ (z-Cosmography I/II/III) and $y = \frac{z}{1+z}$ (y-Cosmography I/II/III) series, and, in all the cases, the results are largely compatible.
Our MCMC method allowed us to obtain constraints on parameter estimation, in particular for higher order cosmographic parameters (the jerk and the snap). It turns out that the deceleration parameter clearly confirms the present acceleration phase; both the estimation of the jerk and the DE parameters, reflect the possibility of a deviation from the $\Lambda$CDM cosmological model. In particular from  the Cosmography II (which combines the SNIa HD, the \textit{ Amati} Gamma Ray Bursts HD with BAO and $H(z)$  data sets  we obtain for the parameter $j_0 \in ( $0.104, 0.92$ )$ at $2\,\sigma$ of confidence, and $j_0 \in ( 0.1, 1.4 )$ at $3\,\sigma$ of confidence. We finally investigate the implications of our results for dark energy: in particular here we focus on the parametrization of the dark energy equation of state (EOS), and we compare the cosmographic and the EOS series. Our analysis indicates that the dark energy equation  of state is evolving for all the parametrizations we considered; moreover the $q(z)$ reconstruction, allowed by our cosmographic analysis, permits a transient acceleration. In a forthcoming paper we are going to compare such indications with a direct and full reconstruction of the EOS from the observational data. We showed that the current data sets are not yet able to discriminate among these alternative scenarios:
the selection of a really high redshift standard rulers is what would really improve the knowledge of the expansion history of our
universe.

\subsection*{Acknowledgments}

This paper was supported in part by the Polish Ministry of Science and Higher Education grant NN202-091839.

\end{document}